\documentclass[a4paper,fleqn,usenatbib]{mnras}
\usepackage{newtxtext,newtxmath}
\usepackage{appendix,apptools}
\usepackage{float}
\usepackage{graphicx}	
\usepackage{amsmath}	
\usepackage{amssymb}	
\usepackage{amsfonts}      
\usepackage{gensymb}        
\usepackage{upgreek}
\usepackage{adjustbox}
\usepackage{subfigure}
\usepackage[bottom]{footmisc}
\usepackage{multirow}
\usepackage[flushleft]{threeparttable}
\usepackage{verbatim}

\newcommand{\Hb}{H$\beta$ } 
\newcommand{\Ha}{H$\alpha$ }

\newcommand{\lya}{\mbox{Ly$\alpha$ }}   

\newcommand{\be}{\begin{equation}}
\newcommand{\ee}{\end{equation}}
\newcommand{\ba}{\begin{eqnarray}}
\newcommand{\ea}{\end{eqnarray}}
\def\hmf{\frac{dn}{dM}}

\begin{document}
\title[Observing the Universe in H$\alpha$ emission]{Tomographic Intensity Mapping versus Galaxy Surveys: Observing the Universe in H$\alpha$ emission with new generation instruments} 

\author[M. B. Silva et al.]{Marta B. Silva$^1$ \thanks{E-mail:
silva@astro.rug.nl}, Saleem Zaroubi$^{1,2,3}$, Robin Kooistra$^1$,  Asantha Cooray$^4$\\
$^{1}$Kapteyn Astronomical Institute, University of Groningen, Landleven 12, 9747AD Groningen, the Netherlands\\
$^{2}$Department of Natural Sciences, The Open University of Israel, 1 University Road, P.O. Box 808, Ra'anana 4353701, Israel\\
$^{3}$Department of Physics, The Technion, Haifa 32000, Israel\\
$^{4}$ Department of Physics $\&$ Astronomy, University of California, Irvine, CA 92697
}

\date{Accepted XXX. Received YYY; in original form ZZZ}



\maketitle

\begin{abstract}
The H$\alpha$ line emission is an important probe for a number of fundamental quantities in galaxies, including their number density, star formation rate 
(SFR) and overall gas content. 
A new generation of low-resolution intensity mapping probes, e.g. SPHEREx and CDIM, will observe galaxies in \Ha emission over a large fraction of the sky 
from the local Universe till a redshift of $z\sim6\, {\rm to}\, 10$, respectively. This will also 
be the target line for observations by the high-resolution Euclid and WFIRST instruments in the $z\sim0.7-2$ redshift range. 
In this paper, we estimate the intensity and power spectra of the \Ha line in the $z\sim0-5$ redshift range using observed line luminosity functions (LFs), 
when possible, and simulations, otherwise. 
We estimate the significance of our predictions by accounting 
for the modelling uncertainties (e.g. SFR, extinction, etc.) and observational contamination. 
We find that Intensity Mapping (IM) surveys can make a statistical detection of the full \Ha emission between $z\sim 0.8-5$.  Moreover, we find that 
the high-frequency resolution and the sensitivity of the planned CDIM surveys allow for the separation of \Ha emission from several interloping lines. We explore 
ways to use the combination of these line intensities to probe galaxy properties.
As expected, our study indicates that galaxy surveys will only detect bright galaxies that contribute up to a few percent of the overall
\Ha intensity. However, these surveys will provide important constraints on the 
high end of the \Ha LF and put strong constraints on the AGN LF. 
\end{abstract}




\section{Introduction}

The H$\alpha$ line is frequently used to detect star-forming galaxies and quasars at low to medium redshifts ($z \lesssim 2.3$). Thanks to
its weak  metallicity dependence, its relatively high luminosity and favorable observing frequency (relative to 
the Ly$\alpha$ line), H$\alpha$ is the main target line of several new space instruments. Due to their very large survey volumes, this new generation of instruments will revolutionize the study of the global properties of galaxy 
formation and evolution and put very stringent constraints on dark energy models \citep{2015Gehrels,2013Amendola}. 

Being a hydrogen recombination line, H$\alpha$ emission is dominated by massive, short-lived stars, such as type O stars 
and early type B stars, characterized by timescales ${\rm < 10-20\, Myr}$.  Hence, it is a good tracer of the instantaneous SFR 
\citep{2009Lee,2015McQuinn}. To constrain the H$\alpha$ intensity, a proper sampling of the high-mass 
end of the initial mass function is required to overcome cosmic variance.
In addition, accounting for medium luminosity  
systems, which also contribute significantly to the overall line intensity requires both large volumes and relatively high 
sensitivities \citep{2015Sobral,2016Pozzetti}.

At low redshift, $z \lesssim 2.3$, the dust attenuation of H$\alpha$ flux is of the order of 1 mag  \citep{2013Sobral,2016Sobral.Stroe}. 
However, it is expected to decline at higher redshifts given the lower dust content of the interstellar medium (ISM).  Furthermore, 
both the observed and the intrinsic \Ha line emission correlate well with the overall galaxy luminosity. In contrast, other metal UV lines, 
such as [OIII]500.7 nm or [OII]372.7 nm depend strongly on the metallicity and dust content of the galaxy \citep{2015Khostovan}.

In recent years, a method for probing gas in
the Universe, called Intensity Mapping, has been proposed \citep{1997madau.Meiksin}. 
The basic idea is to observe the intensity of a specific line emission, e.g. H\,I 21 cm radiation \citep{1997madau.Meiksin, chang10, masui13}, 
and map it at every point in space and redshift, down to  the resolution and sensitivity of the telescope.
Since this method probes a certain spectral line, the redshift of the observed gas parcel comes for free.
Therefore, the 3D mapping made available by this method contains an enormous 
amount of information about the integrated galaxy and IGM emission from each voxel. This new technique is expected to provide a wealth of information  
that is not yet available to current probes. For example, using IM surveys in combination with galaxy surveys  will link the distribution of 
galaxies with IGM overdensities. Given its simplicity and 
many advantages, IM has been extended to several other atomic and molecular emission lines, such as CO, CII and 
Ly$\alpha$, used to probe the  Epoch of Reionization (EOR) \citep{Visbal:2010rz,2011Lidz,2012GongCII,2013Silva,2015Silva}. 
Moreover, IM has also been proposed as a probe of the medium to low redshift Universe in several lines 
\citep{2014Pullen,2014Uzgil,2016Silva,2017Fonseca}. 

Recently, two \Ha IM instruments have been proposed. The first is SPHEREx (the 
Spectrophotometer for the History of the Universe, Epoch of Reionization, and Ice Explorer), which is a NASA Medium-Class Explorer mission selected for phase 
A study in 2017 \citep{2014Dore,2016Dore}. The second instrument is CDIM (the Cosmic Dawn Intensity Mapper \citep{2016CoorayCDIM}) space telescope. \citet{2017Fonseca} have explored the use of the surveys of these two instruments to constrain cosmological models. They found that these missions are able to constrain 
the \Ha power spectrum at $k>0.02\, {\rm h/Mpc}$ down to the few percent accuracy. This makes them useful for probing 
the baryonic acoustic oscillations (BAO) scale. Furthermore, \citet{2017Gong} has demonstrated the 
potential of SPHEREx to put strong constraints
on the star formation rate density (SFRD) for $z\lesssim5$, assuming that it scales linearly with the observed \Ha intensity and 
overlooking the contamination by background lines in the \Ha intensity maps.

Two other interesting instruments for \Ha studies are the Euclid \citep{2011Laureijs} and  WFIRST (Wide-Field Infrared Survey Telescope \citep{2012Green,2015Spergel}) space telescopes. This new generation of instruments will carry out galaxy 
surveys in a very wide area, particularly compared to ground-based instruments. Their spectroscopic 
capabilities will provide data with high-frequency resolution to allow the distinction between two lines with 
small frequency separation, hence significantly reducing line contamination \citep{2016Pozzetti}. 
Although their spectroscopic sensitivity is low, these telescopes will have high photometric sensitivity that will 
allow them to probe down to faint sources -- fainter than the best available ground-based instruments.

We note here that these two instruments will not carry out IM surveys.
Furthermore, the wide-band filters used by Euclid and WFIRST will result in a poor 
determination of the emitted line and thus of the galaxy redshift. Spectroscopic follow-up of the photometric sources 
will only be possible for a few cases and usually for relatively bright galaxies. Another option would be to use 
IM surveys, covering the same area of the sky, to provide better redshift estimates for these galaxies.

In this paper, we study the \Ha emission at $z\lesssim 5$ expected to be observed by these four instruments. We improve upon previous predictions by estimating the uncertainty in the
intensity of the intrinsic \Ha signal and of the dust extinction suffered by \Ha photons. We also
account for the  contribution of quasars to the overall H$\alpha$ luminosity density. Moreover, we explore  several 
possible constraints on astrophysical quantities obtained from the surveys targeting \Ha emission.   

We investigate the contamination of the relevant background/foreground lines in \Ha intensity maps at each redshift. These interloping 
lines include the hydrogen H$\beta$ (486.1 nm) and \lya (121.6 nm) lines, the [SII] (671.7 nm)(673.1 nm) doublet, the [NII] (658.3 nm)(654.8 nm) 
doublet and the ionized oxygen [OII] (372.7 nm) and [OIII] (500.7 nm) lines. 
By doing this study using simulations, we are able to self-consistently explore the foreground removal technique needed to recover the \Ha signal from 
observational intensity maps.

The paper is organized as follows. In Section~\ref{sec:Ha_physics}, we present the physical processes that give rise to the observed \Ha line luminosity in galaxies. 
In Section~\ref{sec:Ha_constraints} we present and discuss current constraints on the galaxy \Ha LF, and on the SFR-halo mass relation. 
The simulation code we run in order to predict the intensity and spatial fluctuations of \Ha emission from both 
galaxies and the IGM is presented in Section~\ref{sec:Simulations}. We follow by describing the four instruments and their planned  \Ha survey characteristics in Section~\ref{sec:Surveys}. In Section~\ref{sec:LineContam} and Section~\ref{sec:contamination4} 
we discuss foreground removal strategies in \Ha intensity maps in the context of these instruments. A comparison 
between constraints from galaxy and IM surveys is made in Section~\ref{sec:Constraints}. The final 
conclusions are presented in Section~\ref{sec:Summary}.

Throughout this paper we assume the best fit cosmological parameters from Planck + WMAP \citep{2014Ade} 
($\Omega_b h^2=0.022032$, $\Omega_m=0.3089$, $h=0.6704$, $Y_P=0.2477$, $n_s=0.9619$ and $\sigma_8=0.8347$). 

\section{H-alpha physics: From SFR to line luminosity}
\label{sec:Ha_physics}

The \Ha line is a Balmer line that corresponds to a transition between energy levels $n = 3$ to $n = 2$ of neutral hydrogen. It is predominantly emitted during 
hydrogen recombinations, but  can also arise due to collisional excitation of this transition. The latter process is mainly relevant 
for warm and neutral gas, such as the boundary region between ionized and neutral gas. Therefore, recombination emission usually dominates the overall \Ha emission in galaxies \citep{2008Cantalupo}. 

It is commonly assumed that the volume average escape fraction of ionizing photons, from galaxies, is very small. This quantity 
is poorly constrained from observations. However, in the few cases where it has been
measured (along a few lines of sight), it is found to be below the 10\% level \citep{2016Vasei}. 
The escape fraction of ionizing photons is highly dependent 
on the gas conditions in the ISM. It can thus range from almost zero up to a few tens of percents. High values of escape fraction 
are a signal that the ISM contains low column density channels along which ionizing photons can easily escape. Given the 
current results from simulations and observations, in the relevant redshift range, escape fractions above $10 - 20$\% are unlikely \citep{2011Boutsia,2014Yajima}.  
As a first approximation, and given that this parameter is degenerate with the galaxy SFR, we will assume that it is zero.  The 
number of hydrogen atom ionizations can then be inferred from the stellar emission spectrum of the galaxy. It should be noted that a zero escape 
fraction is the common assumption in the estimation of the SFRD from observations of nebular emission lines \citep{2008Geach,2016Suzuki}.

Following \cite{1998KennicuttApj}, we connect the SFR to the galaxy luminosity ($L_{\nu}$) in the 1500-2800 \r{A} 
wavelength range by averaging over a \cite{1955Salpeter} initial mass function (IMF) with solar metallicity and with mass 
limits 0.1 to 100 ${\rm M_{\odot}}$.  This gives
\be
SFR({\rm M_{\odot}\, yr^{-1}})=1.4 \times 10^{-28}\, L_{\nu}\, ({\rm ergs\, s^{-1}\, Hz^{-1}}). 
\label{eq:sfr_vs_Lnu}
\ee
Taking the population synthesis galaxy spectra from \cite{1993Bruzual.Charlot} we relate the luminosity density 
at 1500 \r{A} to the rate of hydrogen ionizing  photon emission ($Q_{\rm H}$). This yields the following relation:
\be
SFR({\rm M_{\odot}\, yr^{-1}})=1.08\times 10^{-53}\,Q_{\rm H}\, ({\rm s^{-1}}),
\ee
which is valid for star formation in the age interval $0.1-1$ Gyr, dictated by the assumed IMF.

The timescale for hydrogen ionization is of the order of a few years and the timescale for recombination in the dense 
($n\sim10^2-10^4 {\rm cm^{-3}}$) and ionized ISM is of the order of few hundred years. From a cosmological point of view, 
these are instantaneous processes. Hence, one can safely assume ionization-recombination equilibrium.

For a case B recombination coefficient (the choice of recombination coefficient has little impact on this result), 
an average gas temperature of $10^4\, {\rm K}$ will result in the emission of $\sim 0.45$ \Ha photons per hydrogen 
recombination \citep{1989Osterbrock,1996Madau.Ferguson}. 
This results in the commonly used relation between SFR and the intrinsic \Ha luminosity
\citep{1998KennicuttAraa}:
\be
L^{\rm int}_{\rm H\alpha}\, ({\rm erg\, s^{-1} })=1.26 \times 10^{41} SFR\, ({\rm M_{\odot}\, yr^{-1}}).
\label{eq:Lum_Ha}
\ee
Ideally, one should choose the IMF according to the target population. However, this information is usually not available, 
which introduces IMF related uncertainty into our calculations. For example, using a 
Kroupa IMF \citep{2003Kroupa.Weidner} 
or a Chabrier IMF \citep{2003Chabrier} would result in an increase of $\sim 1.54$ to $1.64$ in 
the $L^{\rm int}_{\rm H\alpha}/SFR$ ratio compared to the Salpeter IMF. Additional uncertainty in this relation
arises from the choice of population synthesis galaxy spectra \citep{2009Lee}. 

The observed \Ha luminosity is obtained by correcting the intrinsic luminosity for dust extinction. This extinction is usually taken 
to be of the order of $A_{\rm H\alpha}=$1 mag, defined as:
\be
L^{\rm obs}_{\rm H\alpha} = L^{\rm int}_{\rm H\alpha}\times 10^{-A_{\rm H\alpha}/2.5}.
\ee 
Dust extinction increases with stellar mass and environment density \citep{2016Sobral.Stroe}. The amount of dust, and therefore 
its line extinction power, correlates with metallicity and is expected to decrease with increasing redshift.

The steep increase of the extinction with stellar mass has a strong impact on the overall extinction, affecting surveys with different flux sensitivity. 
The commonly used extinction value, $A_{\rm H\alpha}\, =\, 1\, {\rm mag}$, corresponds to galaxies with stellar masses 
of a few times $10^{10}\, {\rm M}_{\odot}$ \citep{2016Sobral.Stroe}. From semi-analytical studies \citep{2016Mitchell}, the dark matter 
halo masses corresponding to these stellar masses straddle two orders of magnitude, centered at $M_{\rm halo}\sim 10^{12}\, {\rm M}_{\odot}$. Due 
to their sensitivity limits, Euclid and WFIRST spectroscopic surveys will only be able to observe these luminous and massive galaxies.  Therefore, 
for these surveys the extinction value of $A_{\rm H\alpha}$ = 1 mag is appropriate. However, for surveys capable of probing low luminosity galaxies, 
the overall extinction might be smaller.

Note that, \Ha emission will also suffer extinction due to the dust in the Milky Way. This decrement in the \Ha flux can be estimated with dust maps of the Milky Way. As a reference, corrections due to interstellar extinction in the COSMOS field, in the relevant frequency bands for this study, are of the order of $\lesssim 0.05$ mag \citep{2007ApJS..172...99C}. Moreover, intensity maps will need to be corrected for continuum galactic dust emission and zodiacal light \citep{1998ApJ...500..525S}. 

Unlike Ly$\alpha$, \Ha photons are not efficiently absorbed by neutral hydrogen. Therefore, their flux is expected to suffer 
little to no attenuation due to scattering or dust extinction along their path through to the IGM. 

\section{Modelling H-alpha emission} 
\label{sec:Ha_constraints}

\subsection{H-alpha constraints from observational LFs} 

We make use of the \Ha LFs compiled by \citet{2016Pozzetti}. These include data from the ground-based imaging 
Hi-Z Emission Line Survey (HiZELS) with UKIRT, Subaru and VLT \citep{2013Sobral}, the WISP slitless space-based spectroscopic
survey \citep{2013Colbert} with the Wide Field Camera 3 on HST (HST+WFC3) and from the HST Near Infrared Camera and 
Multi-Object Spectrograph (HST-NICMOS) \citep{2009Shim,2000Hopkins,1999Yan}. 

For simplicity, we fit the \Ha LF using a Schechter fitting function \citep{1976Schechter}:
\be\label{eq:LF}
\Phi(L)dL = \phi_*\left( \frac{L}{L_*} \right)^{\alpha} {\rm exp}\left(-\frac{L}{L_*}\right)\frac{dL}{L_*},
\ee
where $\phi_*$ is a normalization factor with units of inverse volume, $L_*$ is the characteristic luminosity at which there is a break in 
the luminosity function and $\alpha$ is the faint-end slope. This formula usually applies to the intrinsic continuum luminosity of galaxies. 
However, it has also been shown to be a good fit to galaxy line luminosities. In the case of the \Ha line LF, it reproduces the emission from star-forming galaxies well, but 
not the additional contribution from AGN. The observed line luminosity function is then better described by a 
Schechter fitting function plus a power law \citep{2017Matthee}. 

Alternatively, these observations can be fitted to other functions, such as the \cite{1990Saunders} fitting function. 
Therefore, one should bear in mind that the choice of using Schechter fitting function can introduce certain biases to our estimates. 

We evolve the \Ha LF in the redshift range $0<z<2.3$ following the observed LFs mentioned above. Given the small number 
of \Ha flux observations at higher $z$, we infer the \Ha luminosity from observed UV fluxes following the method described in \citet{2016Smit}. 
Basically, this method consists of converting the LFs at 1600\textup{\AA} into SFRs and then to \Ha luminosities using 
Equations~\ref{eq:sfr_vs_Lnu} and \ref{eq:Lum_Ha}, respectively.

 We consider two prescriptions for the \Ha dust extinction. In the first case, we assume an \Ha dust extinction 
of $A_{{\rm H}\alpha}=1$ mag for the entire redshift range.  While in the second case we assume an \Ha dust extinction of 
$A_{{\rm H}\alpha}=1$ mag for the low $z$ ($z\lesssim 2.3$) sample and a
decreasing extinction towards high $z$, reaching $A_{{\rm H}\alpha}=0.475$ mag at $z\sim5$. The latter extinction value is obtained by 
requiring the UV data to fit the \Ha LFs. The high $z$ constraints on the \Ha LF are based on the offset 
between 3.6 $\mu$m fluxes from Spitzer/IRAC and the best spectral energy distribution (SED) fits from the HST upper 
limits and the Spitzer/IRAC photometry \citep{2016Smit}.

The observed low-$z$ \Ha LFs include emission from both star-forming galaxies and AGNs. The percentage of \Ha emitters
powered by AGN activity depends on the luminosity limits.  It can range from a few tens of percent in luminous galaxies to 
practically zero for faint galaxies. For example, \citet{2016Sobral.Kohn} has measured an AGN contribution of $30 \pm 8$\% in the redshift 
range $z \sim 0.8-2.23$ for bright galaxies ($L_{{\rm H}\alpha}>L_{\ast}$). 

The mean intensity of \Ha emission is
\be \label{eq:I_LF}
\bar{I}_{{\rm H}\alpha}(z) = \int_{L_{\rm min}}^{L_{\rm max}} dL \frac{dn}{dL} \frac{L_{{\rm H}\alpha}}{4\pi D_{\rm L}^2}y(z)D_{\rm A}^2, 
\ee
where $D_{\rm L}$ is the proper luminosity distance, $D_{\rm A}$ the comoving angular diameter distance and 
$y(z)[{\rm Mpc\, h^{-1}\, Hz^{-1}}]=d\chi/d\nu$, where $\chi$ is the comoving distance.   
In galaxy surveys, the lower luminosity limit, $L_{\rm min}$, is set by the sensitivity of the instrument. However, the mean intensity due 
to the whole \Ha population would require $L_{\rm min}=0$. The contribution of very faint galaxies, i.e., more 
than two or three orders of magnitude below $L_*$, to the 
total line intensity is quite small for all possible LF shapes. The choice of the upper limit in luminosity for 
the integration, $L_{\rm max}$, should 
be the same for any survey. It is usually taken as the maximum observed luminosity, which is a 
few orders of magnitude above $L_*$ \citep{2016Pozzetti}. 

\begin{figure} 
\begin{centering}  
\includegraphics[angle=0,width=0.5\textwidth]{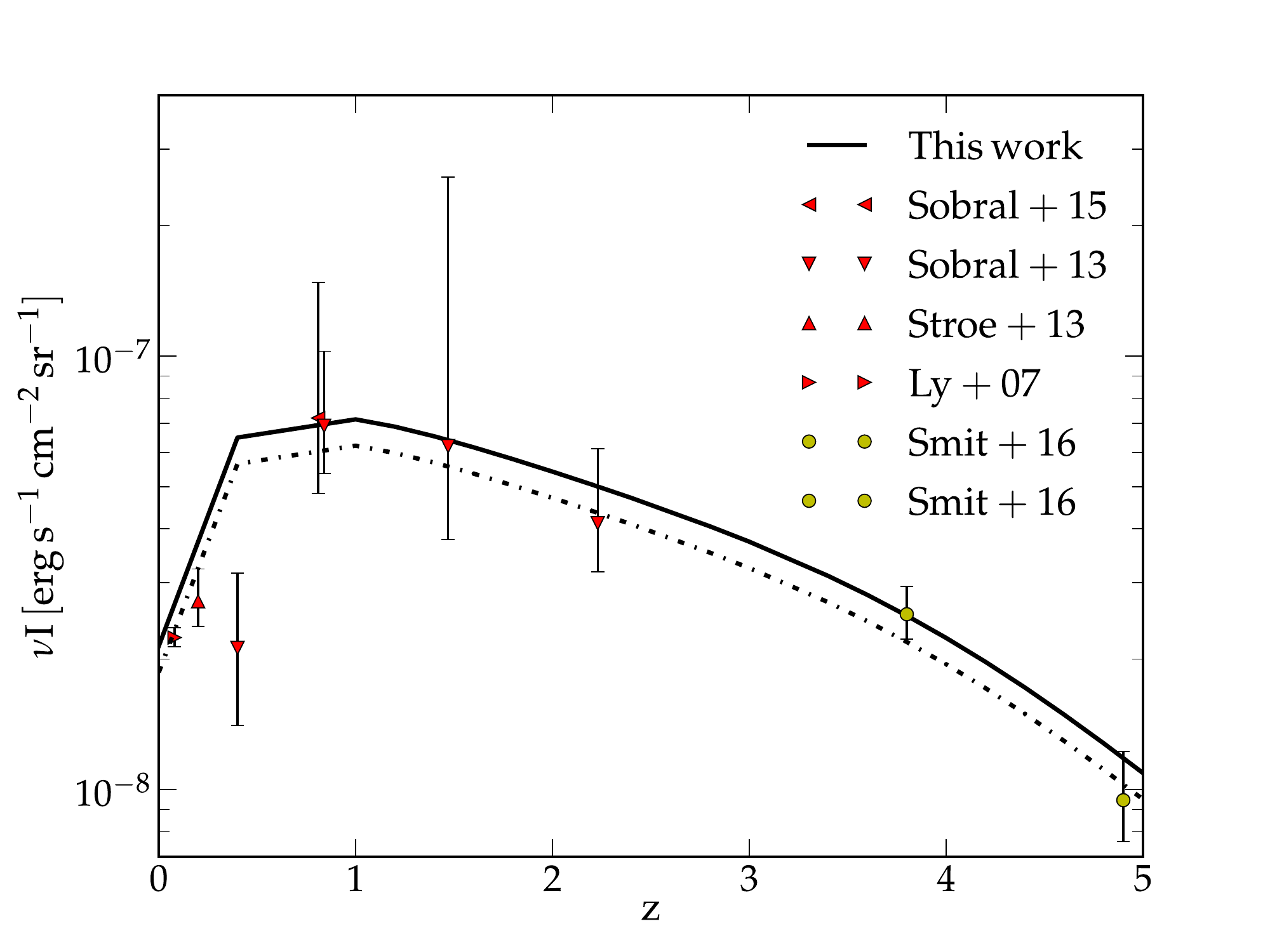}
\caption{Intensity of observed \Ha emission from galaxies. The red triangles correspond to observed \Ha fluxes at $z\lesssim2.3$ 
\citep{1995Gallego,2013Sobral,2015Sobral,2015Stroe}. At higher z, yellow symbols correspond to \Ha intensities obtained 
from the offset between 3.6 $\mu$m fluxes from Spitzer/IRAC and best SED fits from HST upper limits and Spitzer/IRAC photometry, 
following \citet{2016Smit}. The latter correspond to star-forming \Ha emitters only. Also shown is the fit to the observational points based on the \citet{2001Cole} SFRD fitting formula, with (solid line) and without (dotted line) assuming an intensity 
boost of 15\% due to AGN powered \Ha emission. Error bars corresponding to the one-sigma uncertainty on 
the low luminosity end slope of the observational LFs.
}
\label{fig:IHa}
\end{centering}
\end{figure}

The constraints on the mean intensity of \Ha emission from observational LFs are shown as symbols in Figure~\ref{fig:IHa}. Also 
shown is the best fit to the observational data derived from the \citet{2001Cole} SFRD fitting function (black lines). The solid and 
the dashed-dotted lines denote the \Ha intensity with and without a 
15\% boost due to the presence of AGN, respectively \citep{2010Garn,2013Sobral}. The intensity points were derived by integrating the observed 
luminosity functions down to \Ha fluxes of $10^{30}\, {\rm erg\, s^{-1}}$, which is well below the observed flux limit. We fixed this lower 
flux limit and determined the one-sigma error bars from the one-sigma uncertainty in the low luminosity end slope.
Given that this model fits the observational 
constraints well, it will be our base model. The total \Ha intensity (including the AGN contribution) corresponding to this fit is
\ba \label{eq:I_Ha_fit}
\bar{I}_{{\rm H}\alpha}\, (z)\, &=& 2.892\times 10^{-9}\frac{0.027+0.28z}{1+(z/4.8)^{5.3}}\frac{(1+0.15)}{(1+z)^2} \\ \nonumber
&\times& \frac{y(z)}{\rm [Mpc\, h^{-1}\, Hz^{-1}]}\, {\rm erg\, s^{-1}\, cm^{-2}\, sr^{-1}\, Hz^{-1}}.
\ea
We will further discuss this model in Section \ref{sec:Ha_sim}. The uncertainties in the model will be discussed in Section~\ref{sec:Constraints}.

\subsection{H-alpha bias constraints from simulations}
\label{sec:Ha_sim}
We make use of a  SFR model based on simulations, in order to predict \Ha line luminosities where they are not observationally available. This is usually 
the case at higher redshifts and/or at low line luminosities. The use of simulations, which provide a relation between SFR and halo mass, will also allow us 
to estimate the \Ha bias needed to compute the line power spectra.
In this section, we compare how well different analytical functions are able to reproduce our simulated and observational constraints. We choose the best of these fits 
as our base SFR/SFRD model.

\subsubsection{SFR}
\label{subsec:SFR}
\begin{figure}
\begin{centering}  
 \includegraphics[angle=0,width=0.5\textwidth]{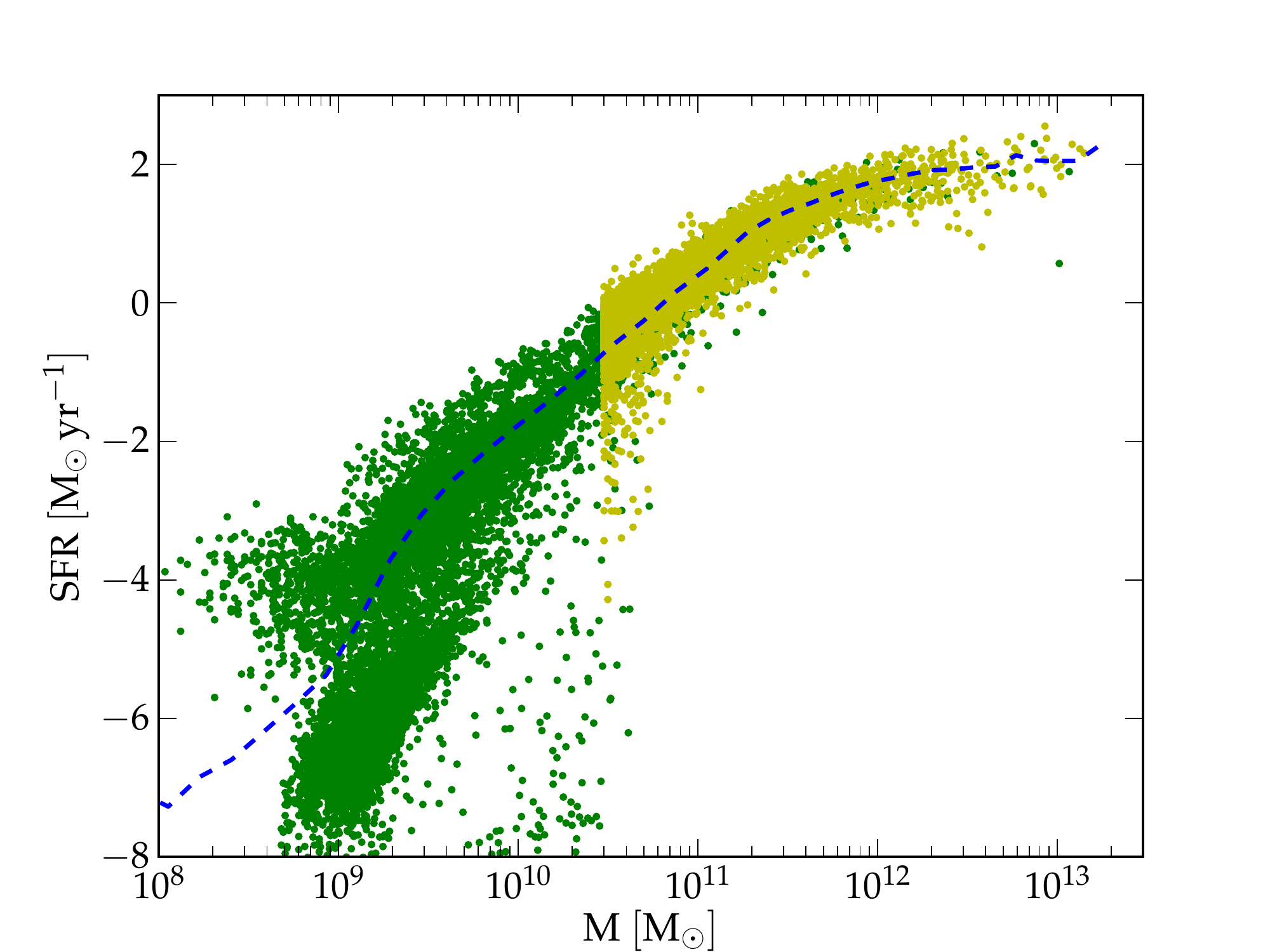}
\caption{
SFR as a function of halo mass at $z\sim2.2$ from the 
\citet{2013Guo.White} galaxy catalog. 
The dashed line marks the average in this relation. The yellow and green dots correspond respectively to DM halos extracted from the 
Millennium and Millennium II cosmological dark matter simulations.
}
\label{fig:sfr}
\end{centering}
\end{figure}

We adopt the catalog of \cite{2013Guo.White} (hereafter, Guo2013), who used a semi-analytic prescription that incorporates astrophysical 
properties of the Sloan Digital Sky Survey (SDSS) to the dark matter halos in the Millennium 
and Millennium II cosmological simulations \citep{2005Springel.White,2009Boylan-Kolchin}. 
This catalog includes an estimate of the galaxy SFR based on its cold gas mass, the fuel for star formation. It is assumed that during a single  
orbital period, 20\% of the cold gas in the galaxy is converted into stars. This is usually presented as the galaxy having an $\epsilon=0.2$ efficiency 
per dynamical cycle of converting cold gas into stars.
The average of the star formation in these galaxies as a function of halo mass can be parameterized with the function
\be
SFR(M)=10^{a}\left(\frac{M}{M_1} \right)^b\left(1+\frac{M}{M_2} \right)^c\, [{\rm M_{\odot}\, yr^{-1}}], 
\label{eq:SFR_Guo}
\ee 
where $M_1=10^8 {\rm M_{\odot}}$. The remaining fitting  parameters for redshifts 
ranging from $z \sim 0-4.8$ can be found in Table \ref{tab:SFR}.
This fit is valid in the $M\sim (10^8-10^{13})\, {\rm M_{\odot}} $ mass range for $z<4$ and in the  
$M\sim (10^8-10^{12})\, {\rm M_{\odot}} $ mass range for $z\gtrsim 4$. At higher halo masses the SFR is assumed to be constant. 
As an example, Figure~\ref{fig:sfr} shows this relation at $z=2.2$. In Section~\ref{subsec:SFRD} we discuss a normalization 
to this relationship that we applied in order to better fit current observational constraints.

\begin{table}
\centering            
\caption{Fitting parameters for the SFR-Halo Mass relation}            
\begin{tabular}{l  c c c c c c }        
\hline\hline                 
     & $z=0$ &  $z=0.8$ & $z=1$  & $z=2.2$ &  $z=4.8$\\    
\hline                 
   a  & $ -9.3 $ &  $-8.75$ & $-8.75$  & $-8.2$   & $-6.7$ \\  
   b  & $  2.6 $ &  $2.7 $  & $2.7$    & $2.7$   &  $2.5$ \\   
   c  & $ -2.9 $ & $-3.0$  & $-3.0$    & $-2.95$  & $-2.6$ \\    
   ${\rm log_{\rm 10}}M_2$ & $ 11.9 $ &  $11.7$  & $11.7$    & $11.6$  & $11.3$ \\ 
  \hline                                  
\end{tabular} 
\label{tab:SFR}     
\end{table}

\subsubsection{SFRD}
\label{subsec:SFRD}
\begin{figure}
\begin{centering}  
\includegraphics[angle=0,width=0.5\textwidth]{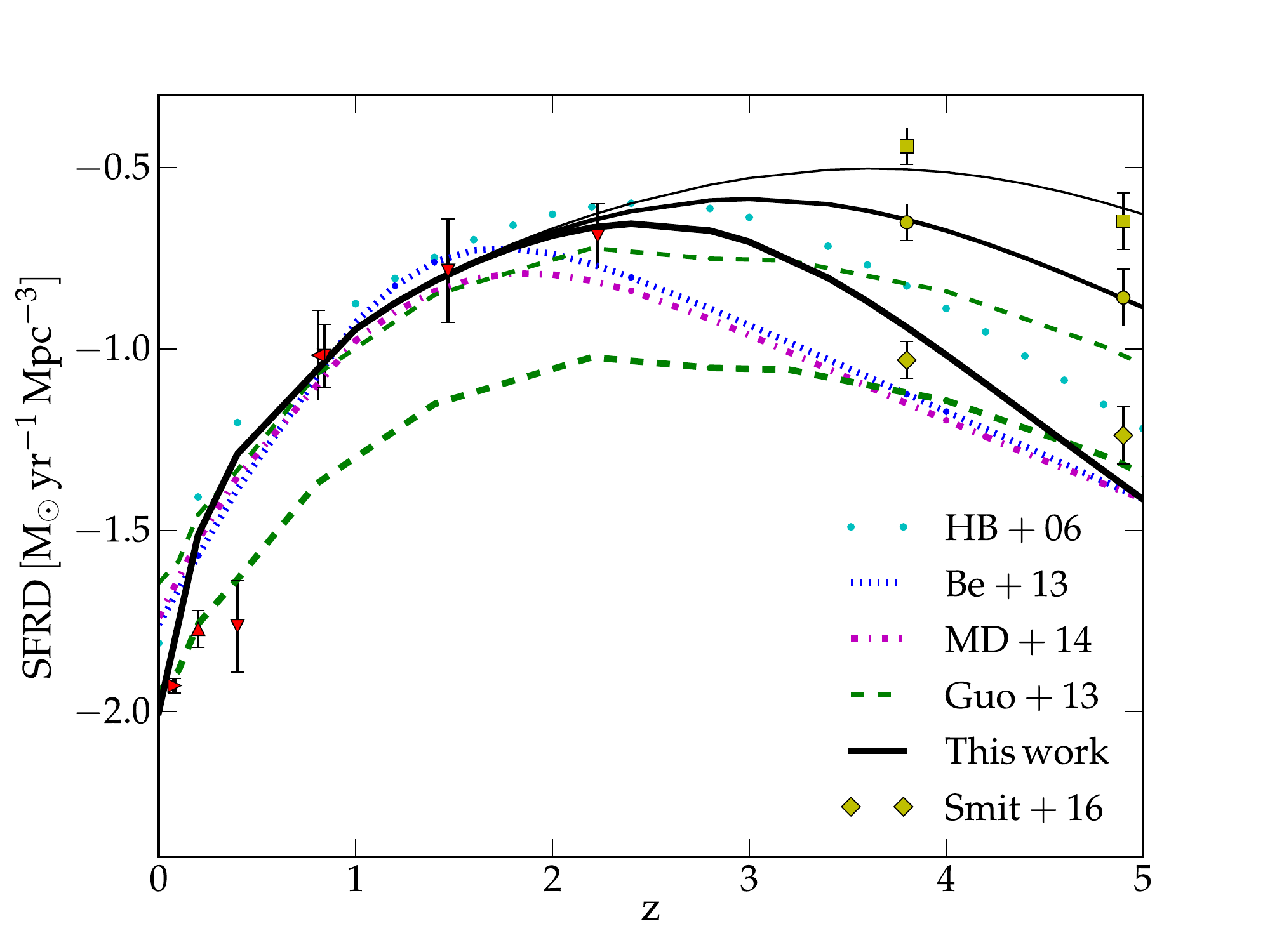}
\caption{
Cosmic SFRD as a function of redshift according to several models. 
The \citet{2013Guo.White} SFRD is shown for the default
efficiency of conversion of gas into stars of $\epsilon=0.2$ (green thick dashed line) and also for $\epsilon=0.4$ (green thin dashed line). 
The symbols correspond to observational constraints based on \Ha emitters. The red triangles are derived from 
\Ha LFs at $z=0.08 $ \citep{2007Ly.Malkan}, at $z=0.0225 $ \citep{1995Gallego}, at $z=(0.4,0.84,1.47,2.23)$ 
\citep{2013Sobral}, at $z=0.81 $ \citep{2015Sobral} and at $z=0.2$ \citep{2015Stroe}. The yellow symbols correspond to 
SFRD constraints, derived from observed \Ha fluxes and UV continuum luminosities \citep{2016Smit}. These symbols assume different values for the extinction
suffered by \Ha emission. Squares, circles and diamonds correspond respectively to: $A_{{\rm H}\alpha}=1.0$, $A_{{\rm H}\alpha}=0.45$ 
and $A_{{\rm H}\alpha}=0.03$. The latter value was obtained assuming an SMC extinction law.}
\label{fig:sfrd}
\end{centering}
\end{figure}

It is worth pointing out the large 
uncertainties in the evolution of the SFRD predicted from simulations, especially towards 
high redshifts. This uncertainty is currently unavoidable due to the poor understanding of 
feedback effects and the large range of halo masses contributing to the SFRD. 
The fitting parameters shown in Table~\ref{tab:SFR} and used in Equation~\ref{eq:SFR_Guo} 
underestimate the SFRD at $z>1$ compared to other observationally based models, such as the 
ones by \citet{2006Hopkins.Beacom} (hereafter, HB+06), \citet{2013Behroozi} (hereafter, Be+13), 
or \citet{2014MadauDickinson} (hereafter, MD+14). 
These models are fits of the SFRD estimated using different sets of observational data in the 
infrared, optical, radio and UV bands.  We compare between these models predictions and the SFRD 
derived from \Ha measurements to illustrate how different probes 
of incomplete data samples at high-$z$ lead to a large uncertainty in the cosmic SFRD evolution. 
For a Salpeter IMF and for the conversion factor between UV luminosity and SFR shown in Equation \ref{eq:sfr_vs_Lnu}, the HB+06 SFRD is
\be
{\rm SFRD}\, (z)\, =\, \frac{h}{0.77}\, \frac{0.012+0.091z}{1+(z/3.3)^{5.3}}\, {\rm M_{\odot}\, yr^{-1}\, Mpc^{-3}},
\ee 
whereas the Be+13 SFRD is
\be
{\rm SFRD}\, (z)\, =\, \frac{0.311}{10^{-0.997(z-z_0)}+10^{0.241(z-z_0)}}\, {\rm M_{\odot}\, yr^{-1}\, Mpc^{-3}},
\ee
with $z_0=1.243$. Finally, the MD+14 SFRD is given by
\be
{\rm SFRD}\, (z)\, =0.015\frac{\left( 1+z \right)^{2.7}}{1+\left[(1+z)/2.9 \right]^{5.6}}\, {\rm M_{\odot}\, yr^{-1}\, Mpc^{-3}}. 
\ee
In Figure~\ref{fig:sfrd} we show the SFRD evolution predicted by the referred models. This figure 
shows that none of these SFRD models provide a good fit to the star formation traced by \Ha emission.

We note that the high redshift points in Figure~\ref{fig:sfrd} are derived from \Ha emitters, which show an excess \Ha flux compared 
to the measured UV fluxes. This might be due to dust correction uncertainties, a bursty or rising star formation history, the shape of 
the ionizing spectrum or other reasons. For a further discussion on this subject see \citet[]{2016Smit}. As a result, at high $z$, 
the SFRs and SFRDs inferred from the \Ha flux using Equation~\ref{eq:Lum_Ha} might be overestimated.  Nevertheless, we will start by 
ignoring this \Ha excess since its origin is not yet clear. In order to be consistent, we will use Equation~\ref{eq:Lum_Ha} to connect SFR 
and \Ha emission for the $z\sim$0-5 redshift range. In Sections~\ref{sec:Summary} and \ref{subsec:Const_SFRD} 
we will discuss the impact of this decision on our conclusions.

 We fit the SFRD traced by \Ha emission by updating the parameters in the \citet{2001Cole} fitting function:
\be
{\rm SFRD}\, (z)\, =\,  \frac{0.01+0.1036z}{1+(z/z_1)^{e}}\, {\rm M_{\odot}\, yr^{-1}\, Mpc^{-3}},
\label{eq:sfrd_model}
\ee
where $z_1=4$ and $e=5$. These parameters can be derived from the \Ha intensity given by Equation \ref{eq:I_Ha_fit}, using 
the conversion factor between SFR and \Ha luminosity in Equation \ref{eq:Lum_Ha}. We assume a dust extinction of 
$A_{{\rm H}\alpha}=1\, {\rm mag}$ at low redshift ($z\lesssim 2.3$) and a
lower extinction of $A_{{\rm H}\alpha}=0.45\, {\rm mag}$ at $z\sim4-5$ (shown as the middle thickness black solid line 
in Figure~\ref{fig:sfrd}). The same fit for a constant \Ha extinction of  $A_{{\rm H}\alpha}=1\, {\rm mag}$
can be obtained using $z_1=4.8$ and $e=5.3$ (shown as the thin black solid line in Figure~\ref{fig:sfrd}).
The SFRD formula for  a Small Magellanic Cloud (SMC) extinction curve \citep{2003Gordon}, can be obtained with $z_1=3.25$ and $e=5.9$ (shown 
as the thick black solid line in Figure~\ref{fig:sfrd}).  

From the current \Ha observations, one can trace the LF and use it to derive the cosmic SFRD. However, in order to estimate the \Ha
intensity power spectrum we need to establish a relation between SFR and DM halo mass. Since we cannot observationally probe the masses of the DM 
halos hosting the \Ha emitters we adopt the SFR-halo mass relations described in Section \ref{subsec:SFR}. We normalize these relations to our 
observationally based SFRD fit given by Equation~\ref{eq:sfrd_model} (with $z_1=4$ and $e=5$) and adopt them as our base SFR model. 

In the following sections, we use this SFR model to estimate the average \Ha flux originating from a DM halo, 
the \Ha bias and finally the power spectra of \Ha intensity spatial fluctuations. Note that at low redshift there are a 
few observational points below our theoretical model.
However, the adopted simple, yet physically based model cannot properly fit these points and the constraints at $z\sim1$ at the same time.
\subsubsection{Power Spectrum}
The total power spectrum of H$\alpha$ emission can be written as 
\be
P_{\rm H\alpha}^{\rm tot}(k,z)=P_{\rm H\alpha}^{\rm clus}(k,z)+P^{\rm shot}_{\rm H\alpha}(z),
\ee
where the first term accounts for the galaxy clustering and is given by:
\be\label{eq:P_Lya} 
P_{\rm H\alpha}^{\rm clus}(k,z) = \bar{b}_{\rm H\alpha}^2 \bar{I}_{\rm H\alpha}^2 P_{\delta \delta}(k,z),
\ee
where $b_{\rm H\alpha}$ is the bias between \Ha emission and the matter power spectrum ($P_{\delta \delta}$). 
The \Ha luminosity weighted bias is
\be
\label{eq:lumbias}
\bar b_{\rm H\alpha} \left( z \right) \equiv\frac{\int^{M_{\rm max}}_{M_{\rm min}}{dM} ~b\left( M,z\right) L_{{\rm H}\alpha}(M,z) ~\hmf }{\int^{M_{\rm max}}_{M_{\rm min}}{dM}~ L_{{\rm H}\alpha}(M,z) ~\hmf }\,,
\ee
where $b\left( M,z\right)$ is the halo bias and $dn/dM$ is the DM halo mass function. The integration limits for the halo mass are 
$M_{\rm min}=10^{\rm 8}{\rm M_{\odot}}$ and $M_{\rm max}=10^{\rm 15}{\rm M_{\odot}}$.
The clustering term dominates the observed line power spectrum at large scales. 
Whereas, on small scales it is dominated by the second term, namely the shot noise caused by 
the discrete distribution of galaxies \cite[e.g.,][]{Visbal:2010rz}:
\be\label{eq:Pshot_Lya}
P^{\rm shot}_{\rm H\alpha}(z) = \int_{L_{\rm min}}^{L_{\rm max}} dL \frac{dn}{dL} \left[\frac{L_{\rm gal}^{\rm H\alpha}}{4\pi D_{\rm L}^2}y(z)D_{\rm A}^2\right]^2.
\ee
The relation between 
line luminosity and halo mass, used in Equation~\ref{eq:lumbias}, has a large scatter. Moreover, it is rarely possible to observationally probe DM halo masses.
The bias of \Ha emitters 
is observationally constrained at $z\sim2.24$ to be $b_{\rm H\alpha}\, =\, 2.4^{+0.1}_{-0.2}$ \citep{2012Geach}. 

This bias was derived by comparing a sample of 370 \Ha emitters detected by HiZELS with predictions from GALFORM semi-analytic models \citep{2000Cole}. 
The HiZELS uniform complete sample contains emitters with luminosities down to $L_{\rm H\alpha} = 2\times 10^{42}\, {\rm erg\, s^{-1}}$. However, the 
referred bias was estimated assuming a lower luminosity limit of $L_{\rm H\alpha}^{\rm min}\, =\, 10^{41}\, {\rm erg\, s^{-1}}$. Since the \Ha bias 
decreases towards lower luminosities, the total \Ha emission bias might be lower than the quoted value.
 
\begin{figure}
\begin{centering}  
\includegraphics[angle=0,width=0.5\textwidth]{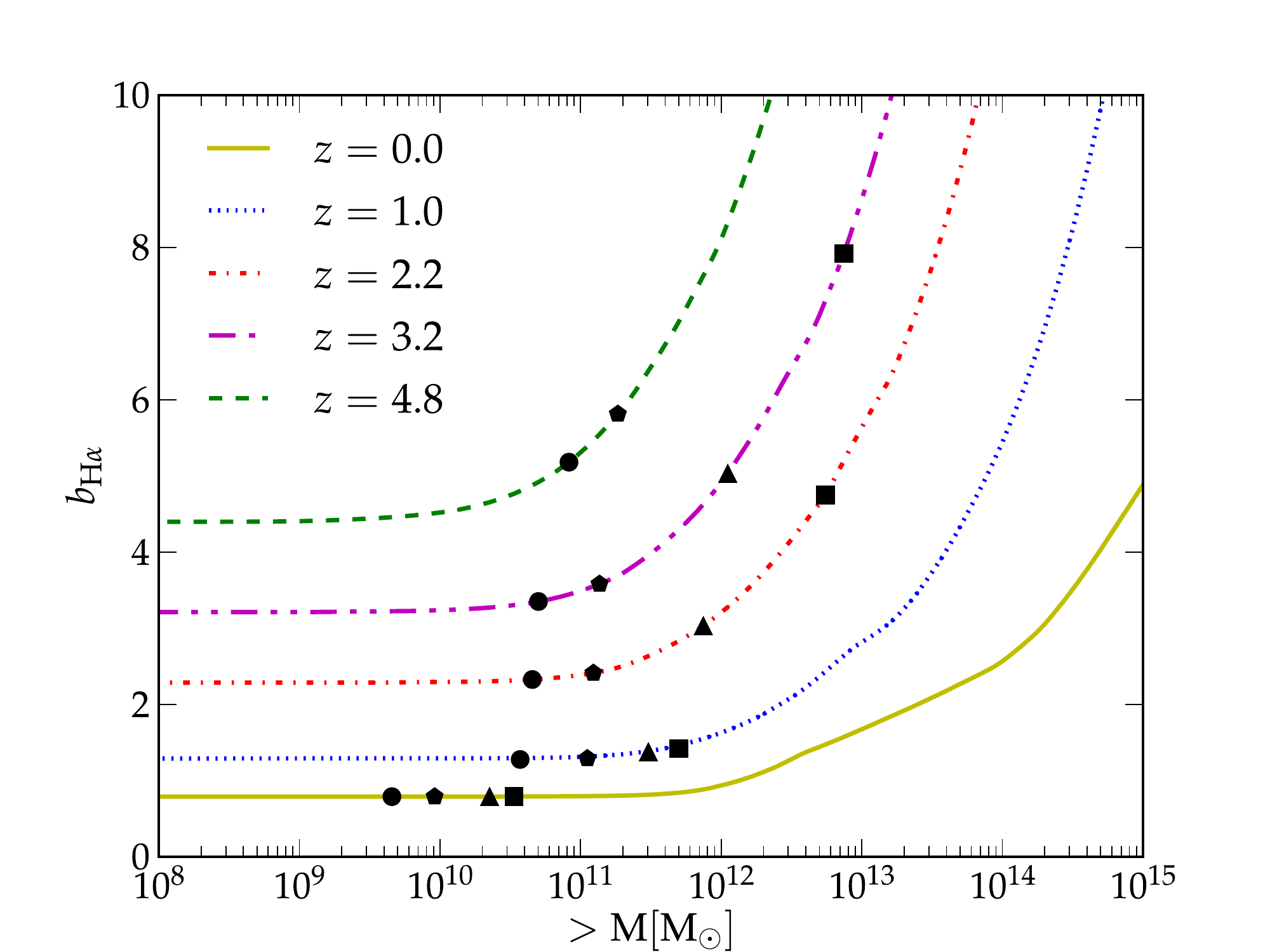}
\caption{Cumulative bias of \Ha emission. The x-axis corresponds to the minimum mass of a halo used on the bias integration. The 
circles, pentagons, triangles, and squares correspond respectively to flux limits of ($10^{-18}$, $10^{-17}$, $1.2\times10^{-16}$, 
$3\times10^{-16}$) ${\rm erg\, s^{-1}\, cm^{-2}}$.}
\label{fig:bias}
\end{centering}
\end{figure}

Given the lack of observational constraints for the full redshift range of interest, we estimate the \Ha bias assuming that \Ha emission scales with the 
galaxy SFR following Equation~\ref{eq:Lum_Ha}. 
The cumulative \Ha bias as a function of redshift is shown in Figure~\ref{fig:bias} using our SFR model. This figure clearly shows that the current observational constraints (limited to high fluxes) cannot be used  to calculate the bias for a wide range of galaxy luminosity, hence a theoretical model is needed.

The total \Ha bias depends on the evolution of the slope of the relation between the SFR and the mass of a DM halo, but it is independent of the amplitude of this 
relation. Nonetheless, we 
use our normalized SFR-halo mass relation described in Sections~\ref{subsec:SFR} and \ref{subsec:SFRD}, so that we can estimate the 
\Ha bias as a function of minimum halo mass. This allows us to estimate the cumulative \Ha bias according to several flux cuts, relevant for the upcoming \Ha surveys.

At low redshifts and for large halo masses each DM halo contains several galaxies. However, there is usually one galaxy dominating the SFR in the halo and 
so, for simplicity, we assume that the \Ha flux in a halo corresponds to the emission from one galaxy.

\section{Simulated H-alpha emission from galaxies and from the IGM} 
\label{sec:Simulations}
Intensity mapping surveys will be sensitive to \Ha emission from galaxies. However, these surveys will also detect \Ha emission 
from the large-scale IGM filaments connecting the DM halos in which galaxies reside. In this section, we use simulations to estimate 
the \Ha emission from these two media 
and compare their relative contribution to the total \Ha emission intensity and power spectrum.

\subsection{The simulation}
\label{subsec:simulated_emission}

The intensity of IGM \Ha emission scales nonlinearly with the gas density and temperature.
Therefore, we run a simulation code with a high spatial resolution in order to model the local properties of the gas.
We start with a dark matter only run made with the parallel 
code Gadget 2 \citep{2001Springel,2005Springel}. This simulation covers a volume of $(200\, {\rm Mpc\, h^{ -1}})^3$ with $1024^3$ particles, each with a mass of $6.5\times10^8\, {\rm M_{\odot}\, h^{-1}}$.

Simulation outputs in the redshift range $\sim0-5$ are used to estimate the \Ha intensity.  In addition, outputs 
up to $z\sim10$ are used to estimate the contamination by background lines in \Ha intensity maps.

The particles are distributed in 3D boxes with $1200^3$ cells following the cloud in cell method. 
In order to model the conditions of the IGM gas we assume that the spatial distribution of the baryonic matter follows that of dark matter. 
The gas temperature, the neutral hydrogen number density ($n_{\rm HI}$), the ionized hydrogen number density ($n_{\rm HII}$) 
and the electron number density ($n_{\rm e}$), 
are estimated following the prescription outlined in \cite{2017Kooistra}.

\begin{figure}
\hspace{-5 mm}
\includegraphics[angle=0,width=0.55\textwidth]{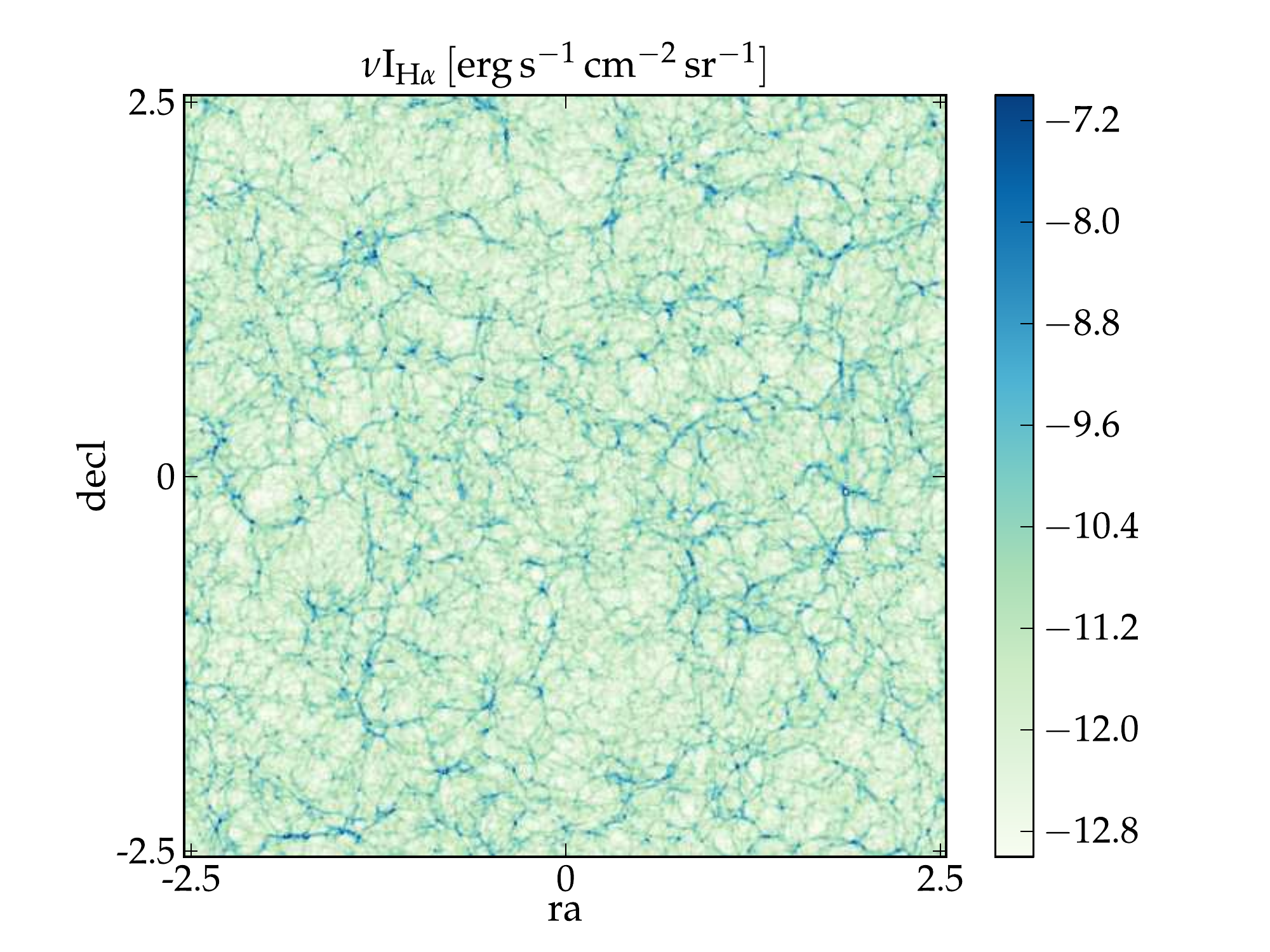}
\caption{Map of \Ha emission from galaxies and from the IGM at $z\sim2$. 
The map scale corresponds to ${\rm log_{10}}\left( \nu {\rm I}_{\rm H\alpha}[{\rm erg\, s^{-1}\, cm^{-2}\, sr^{-1}}] \right)$.}
\label{fig:map_Ha}
\end{figure}

Additionally, the Amiga halo finder code \citep{2004Gill} is used in order to 
extract DM halos from the Gadget 2 outputs. The minimum halo mass in our simulation is $M_{\rm min}= 6.5\times 10^{9}\, {\rm M}_{\odot}$.
To each of these halos we attribute a SFR (normalized to our SFRD model) from a random halo with a similar mass from the Guo2013 galaxy catalog. 
The \Ha emission from the DM halos is obtained with Equation~\ref{eq:Lum_Ha}, assuming a dust extinction of $A_{{\rm H}\alpha}=1\, {\rm mag}$. 
Figure~\ref{fig:map_Ha} shows a map of the total \Ha emission from galaxies and from the IGM at redshift 2 (notice the logarithmic color scale).
This figure clearly shows that the contribution of the diffuse component is subdominant. The procedure used to derive the IGM \Ha emission 
from the simulation is  outlined in Section \ref{subsec:IGM_emission}. 

\subsection{H-alpha IGM emission}
\label{subsec:IGM_emission}

At $z\sim 5$, the IGM gas is kept highly ionized by the ultraviolet background radiation (UVB) together with ionizing radiation from local sources. Most of this gas is located in a 
large scale filamentary structure connecting galaxies and galaxy clusters. The relative overdensity of filamentary gas allows for the existence 
of small clumps of neutral gas. Therefore, in these filaments, there will be \Ha emission originating in hydrogen recombinations and 
collisional excitations.
The luminosity density per comoving volume of \Ha emission from hydrogen recombinations in the IGM is
\be
{\rm \ell}^{\rm IGM}_{\rm rec}(z)= f_{\rm rec}\, \dot{n}_{\rm rec}\,  E_{\rm H\alpha},
\label{eq:Ha_IGM_rec}
\ee
where $E_{\rm H\alpha}=1.89\, {\rm eV}$ is the energy of an \Ha photon. $f_{\rm rec}$ is the probability of emission of an \Ha photon 
per recombination of a hydrogen atom. The value of $f_{\rm rec}$ at a gas temperature  $T\, =\, 10^4\, {\rm K}$, is $0.45$ \citep{2006Osterbrock}.
The number density of recombinations per second, $\dot{n}_{\rm rec}$, is:
\be
\dot{n}_{\rm rec}(z)=\alpha_B(T)\, n_e(z)\, n_{\rm HII}(z).
\label{eq:nrec_s}
\ee
Here $\alpha_B$ is the case B recombination coefficient for hydrogen.
The gas temperature in the IGM can be much lower than that of the ISM. Therefore, we 
also consider a temperature dependent effective recombination coefficient for the \Ha emission, 
$\alpha_B^{\rm H\alpha}(T)=f_{\rm rec}(T)\, \alpha_B(T)$. We take the fitting formula from \citet{2015Raga.Castellanos}, hereafter Raga+15:
\be
\alpha_{\rm B}^{\rm H\alpha}=\frac{4.85\times10^{-23}}{T^{0.568}+3.85\times10^{-5}T^{1.5}}\, {\rm cm^3\, s^{-1}}.
\ee
This fit follows the same trend as the tabulated values from \citet{2006Osterbrock}.
The \Ha luminosity due to collisional excitations is
\be
{\rm \ell}_{\rm coll}^{\rm IGM}=E_{{\rm H}\alpha}\, n_{\rm e}\, n_{\rm HI}\, q_{\rm eff}^{{\rm H}\alpha},
\ee
where $n_{\rm HI}$ is the neutral hydrogen number density. The parameter $q_{\rm eff}^{\rm H\alpha}$ is the effective 
collisional excitation coefficient for \Ha emission, which is taken from Raga+15 as
\be
q_{\rm eff}^{{\rm H}\alpha}=\frac{3.57\times10^{-17}}{T^{0.5}}e^{-140360/T}\left(1+\frac{7.8}{1+5\times10^5/T} \right).
\ee

\begin{figure}
\begin{centering}  
\includegraphics[angle=0,width=0.5\textwidth]{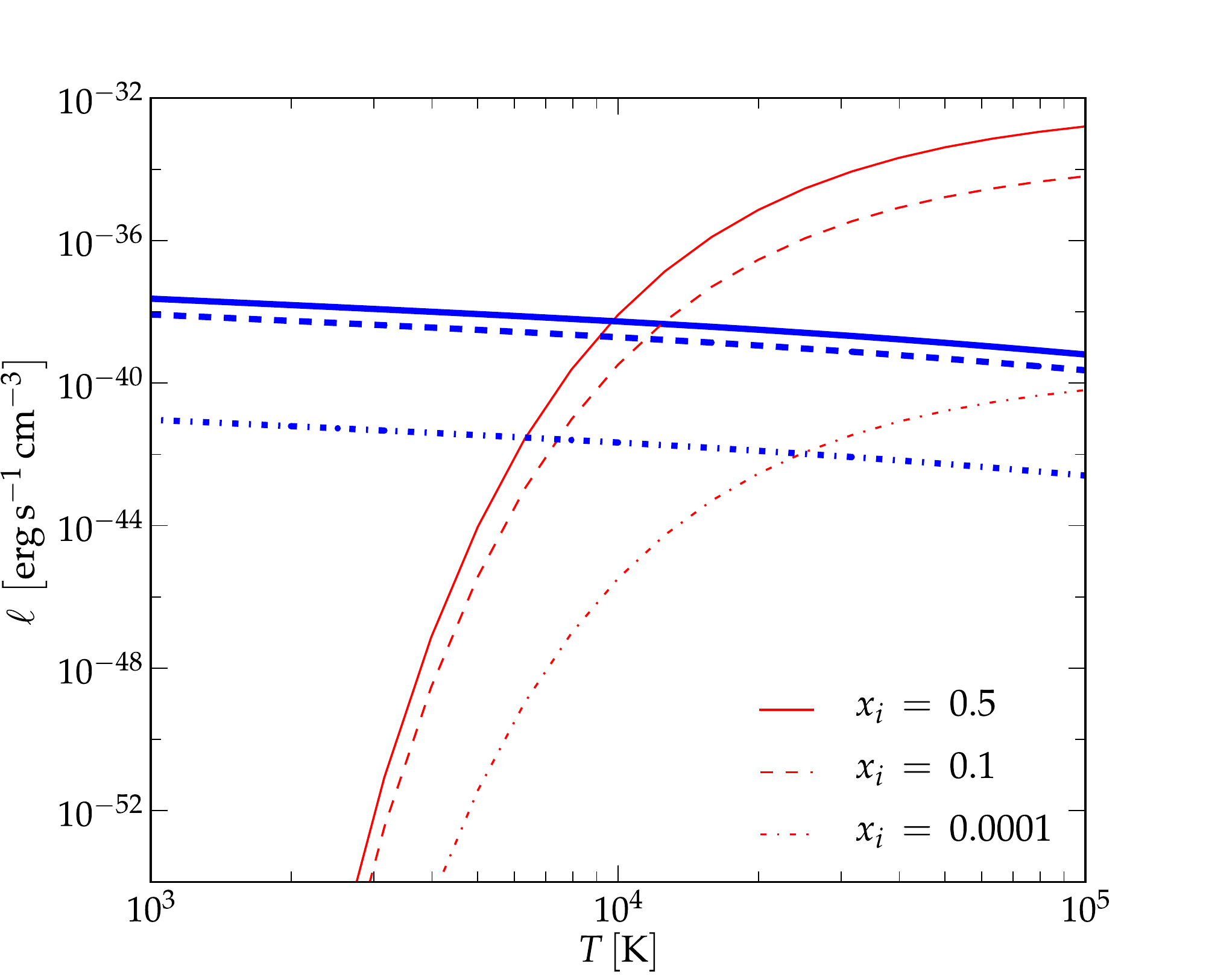}
\caption{Luminosity density of \Ha emission from gas with a hydrogen density of $n_{\rm H}\, =\, 10^{-6}\, {\rm cm^{-3}}$. Recombination 
\Ha emission (blue thick lines) and collisional excitation \Ha emission (red thin lines) are shown for gas with an ionized 
fraction of $x_i\, =\, (0.5\, ,0.1\, ,0.0001)$, for respectively the solid, dashed and dot-dashed lines.
Note, that in the IGM the average hydrogen ionized fraction is very small, of the order of $10^{-5}$ at $z\sim0$.}
\label{fig:coll_rec}
\end{centering}
\end{figure}

The relative importance of recombination and collisional excitation for \Ha emission is illustrated in Figure~\ref{fig:coll_rec}. The 
plotted luminosity densities assume hydrogen number densities of $n_{\rm H}\, =\, 10^{-6}\, {\rm cm^{-3}}$, $n_{\rm HII}=n_{\rm e}=x_i n_{\rm H}$ and 
$n_{\rm HI}=(1-x_i)n_{\rm H}$, where $x_i$ is the gas ionized fraction. As a reference, the average hydrogen number density in the IGM is about ${n_{\rm H}\, =1.9 \times 10^{-7} \rm cm^{-3}}$. The assumed ionized fractions 
of the gas are presented in the figure. 
The gas temperature and its ionization state are both positively correlated with the strength of the extragalactic background radiation. 
The assumption of thermal and ionization equilibrium sets the HI gas temperature to $T\, \sim\, 10^4\, {\rm K}$. Hence, recombinational 
emission is the dominant process for generating \Ha photons in this medium. 

The intensity of \Ha emission from the IGM is
\be
I_{\rm H\alpha}^{\rm IGM}(z)=\frac{\left({\rm \ell}^{\rm IGM}_{\rm rec}+{\rm \ell}^{\rm IGM}_{\rm coll}\right)\, D_A^2}{4\pi D_L^2}y(z).
\ee
Using this equation, we estimate the \Ha intensity for each cell of our simulation boxes.
Cells with densities above the threshold for collapse,
which at $z\sim0$ is $\Delta_{\rm c} \sim 328$, should contain galaxies. These galaxies will contain most of the baryonic mass in the cell. The remaining 
baryonic mass will be highly heated and ionized by 
the local sources. Therefore, we assume that the \Ha IGM emission in these regions is zero. 

\begin{figure}
\begin{centering}  
\includegraphics[angle=0,width=0.5\textwidth]{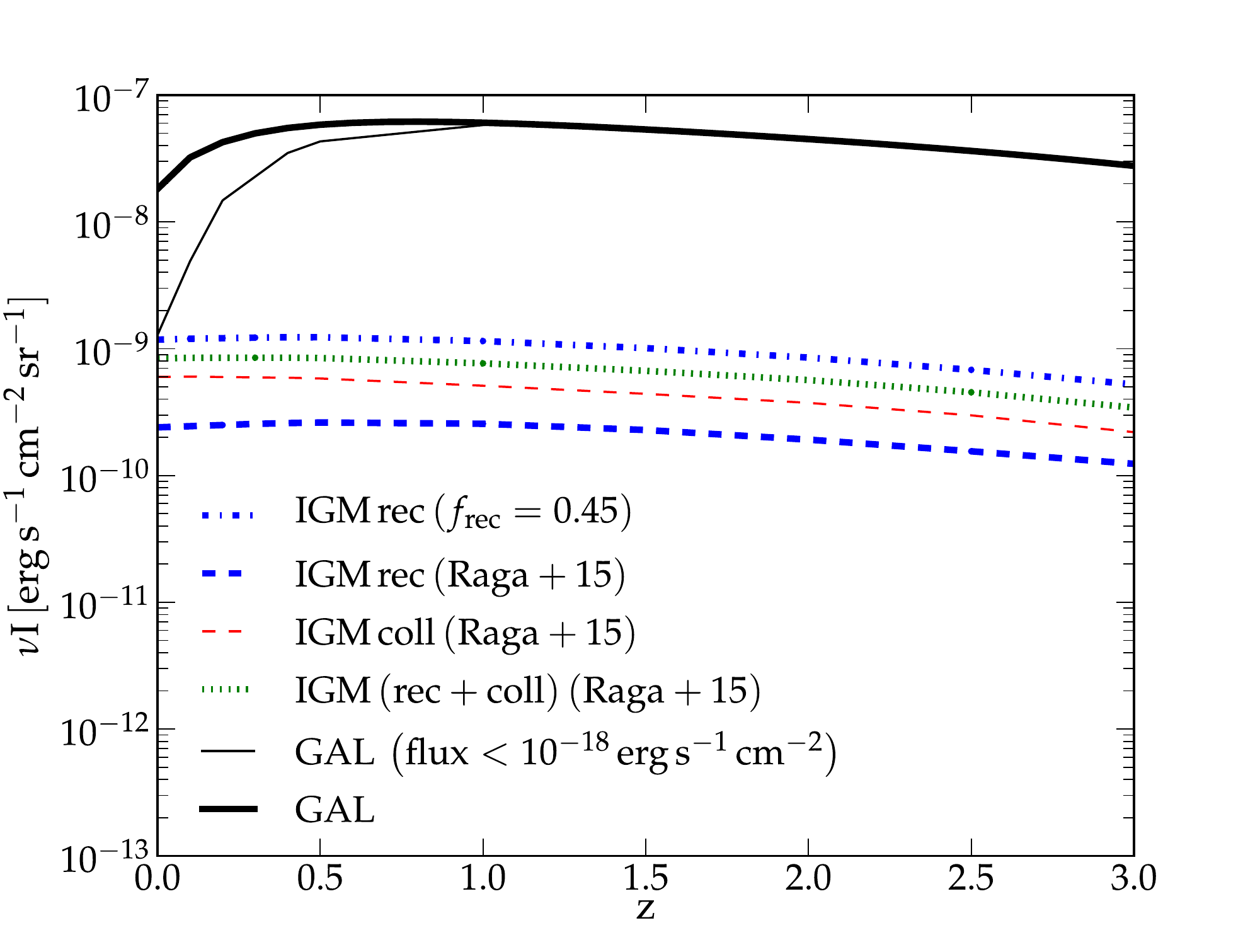}
\caption{Intensity of \Ha emission from galaxies (two top lines) and from the IGM (bottom lines). The black thin solid 
line accounts for the emission from galaxies with \Ha fluxes ${\rm flux}<10^{-18}\, {\rm erg\, s^{-1} cm^{-2}}$. The blue dashed-dotted line 
assumes $f_{\rm rec}\, =\, 0.45$, while the red lines assume the 
temperature dependent \Ha emission rates from Raga+15. 
}
\label{fig:I_Ha_IGM_GAL}
\end{centering}
\end{figure}

Figure~\ref{fig:I_Ha_IGM_GAL} shows the redshift evolution of the simulated \Ha intensity from the IGM and from galaxies. The considered model results in galaxy
\Ha emission dominating over \Ha emission from the IGM. This figure also shows that, after masking the emission from galaxies with \Ha fluxes above 
$10^{-18}\, {\rm erg\, s^{-1}\, cm^{-2}}$, the 
remaining signal is still dominated by emission from low luminosity galaxies and not from IGM emission. At $z\sim0$, the 
intensities of these two faint sources are similar due to the low redshift emission being highly dominated by bright sources.

\begin{table*}
\centering            
\caption{Parameters for the spectroscopic surveys}            
\begin{tabular}{l  c c c c c c }        
\hline\hline                 
Instrument            & $\lambda      $ & $z_{{\rm H}\alpha}$  & $\delta_{\theta}$ & $R$ & Flux limit                       & FOV\, total \\ 
                      & ${\rm (\mu m)}$ &                      & ${\rm (arcsec)} $ &     & ${\rm (erg\, s^{-1}\, cm^{-2})}$ & $({\rm deg^2})$\\
\hline                 
    SPHEREx\,  deep   & $0.75-4.18$     & $0.1428-5.369$       & $6.2$     & $41.4$      & $10^{-17}$             & $200$ \\  
    SPHEREx\,  deep   & $4.18-5.0$      & $5.369-6.6187$       & $6.2$     & $135$       & $10^{-17}$             & $200$ \\  
    CDIM\,  deep      & $0.75-7.5$      & $0.14-10.43$         & $1$  	   & $500$       & ${\rm \le 4 \times 10^{-18}}$ & $25$ \\  
    CDIM              & $0.75-7.5$      & $0.14-10.43$         & $1$  	   & $500$       & ${\rm 10^{-17}}$       & $300$ \\  
    Euclid\, deep     & $1.1-2.0$       & $0.68-2.036$         & $0.3$     & $250$       & ${3\times 10^{-16}}$   & $40$ \\  
    WFIRST            & $1.35-1.95$     & $1.05-1.96$          & $0.15$    & $75$        & ${1.2\times 10^{-16}}$ & $2227$ \\  
  \hline                                  
\end{tabular}
\label{tab:instruments}     
\end{table*}

\subsection{H-alpha emission power spectra}
\label{subsec:Ha_power_spectra}

\begin{figure*}
\vspace{-10pt}
\centerline{
\hspace{2pt}
\resizebox{!}{!}{\includegraphics[angle=0,width=0.35\textwidth]{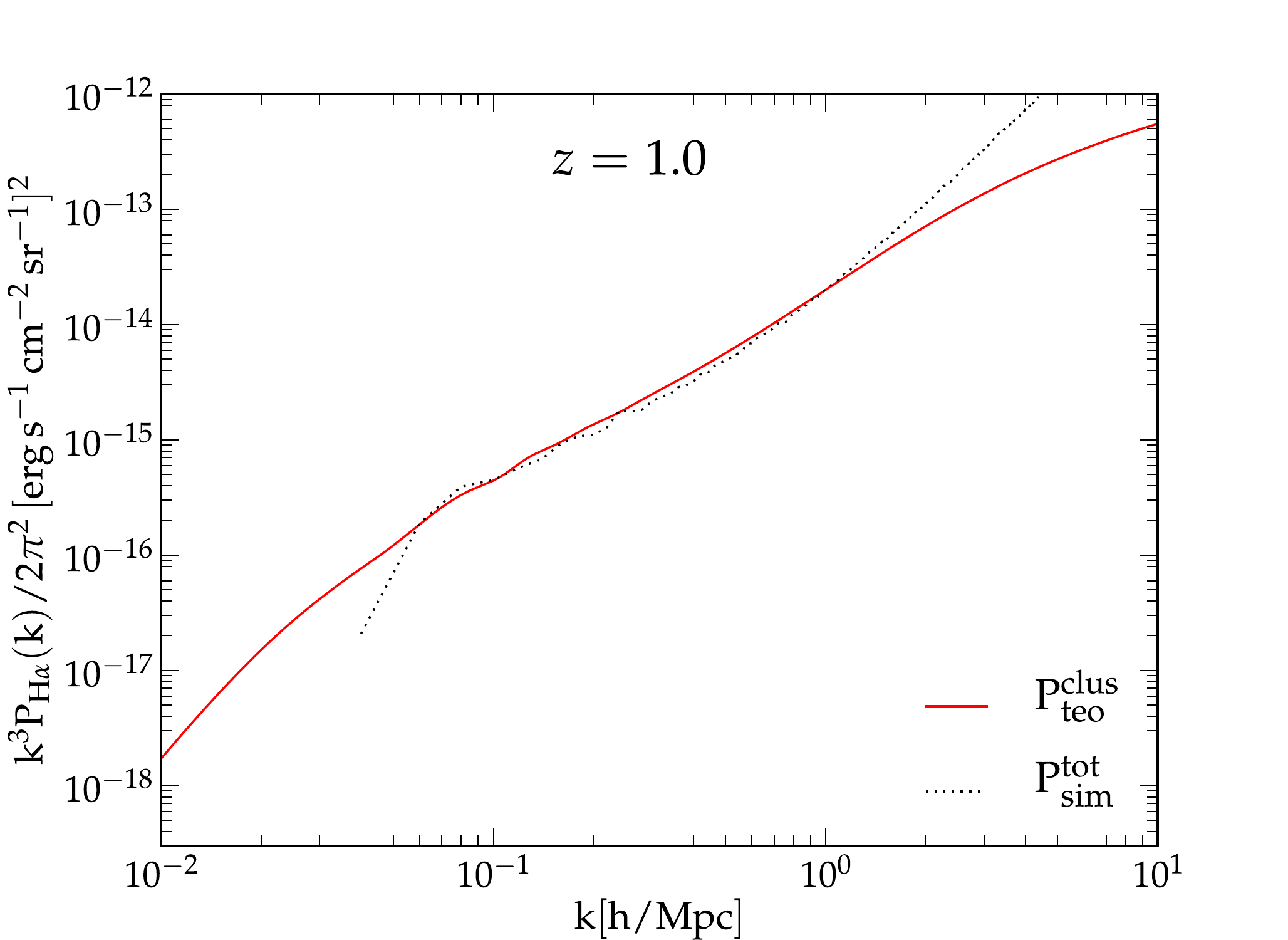}}
\hspace{-12pt}
\resizebox{!}{!}{\includegraphics[angle=0,width=0.35\textwidth]{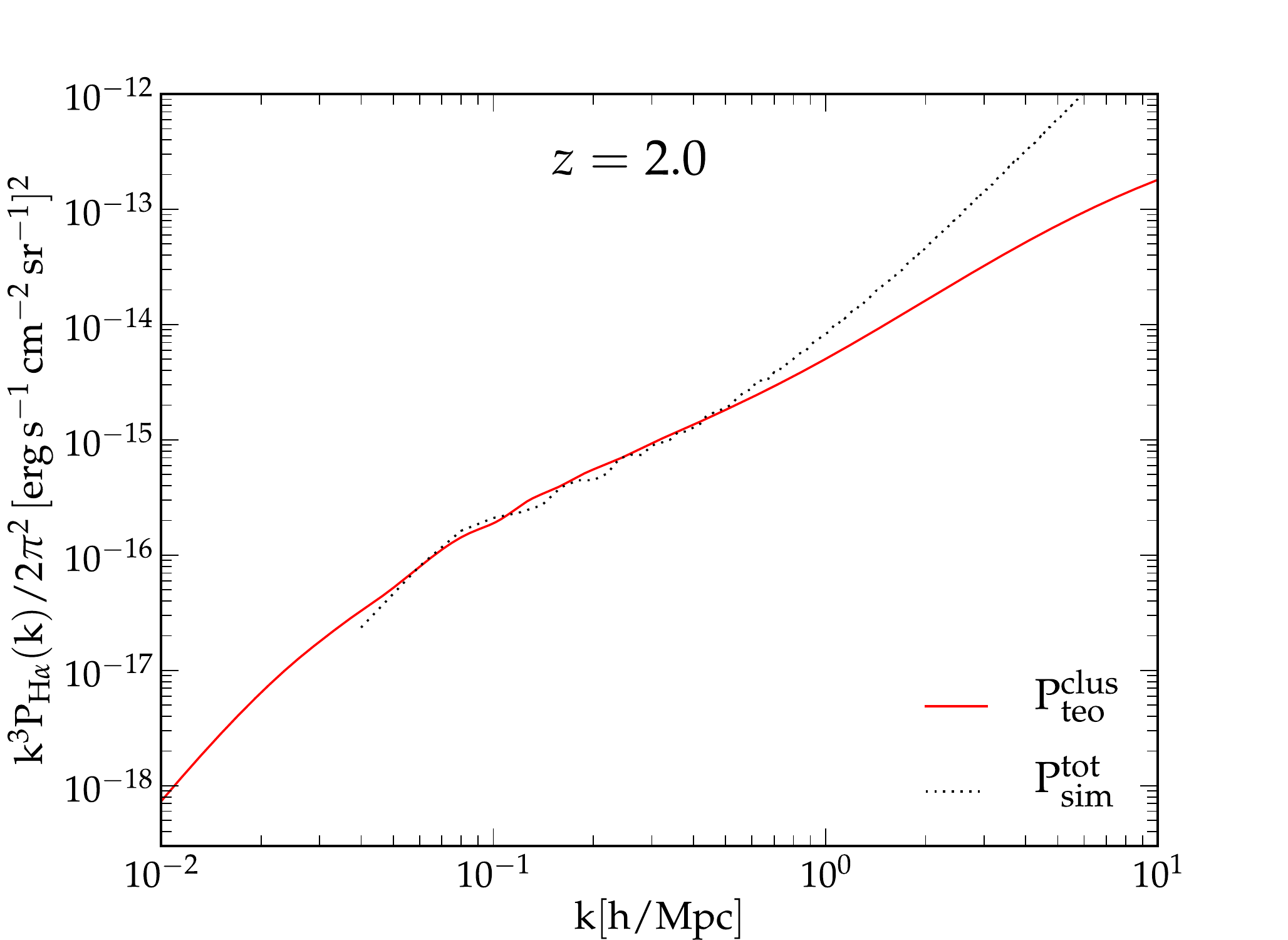}}
\hspace{-12pt}
\resizebox{!}{!}{\includegraphics[angle=0,width=0.35\textwidth]{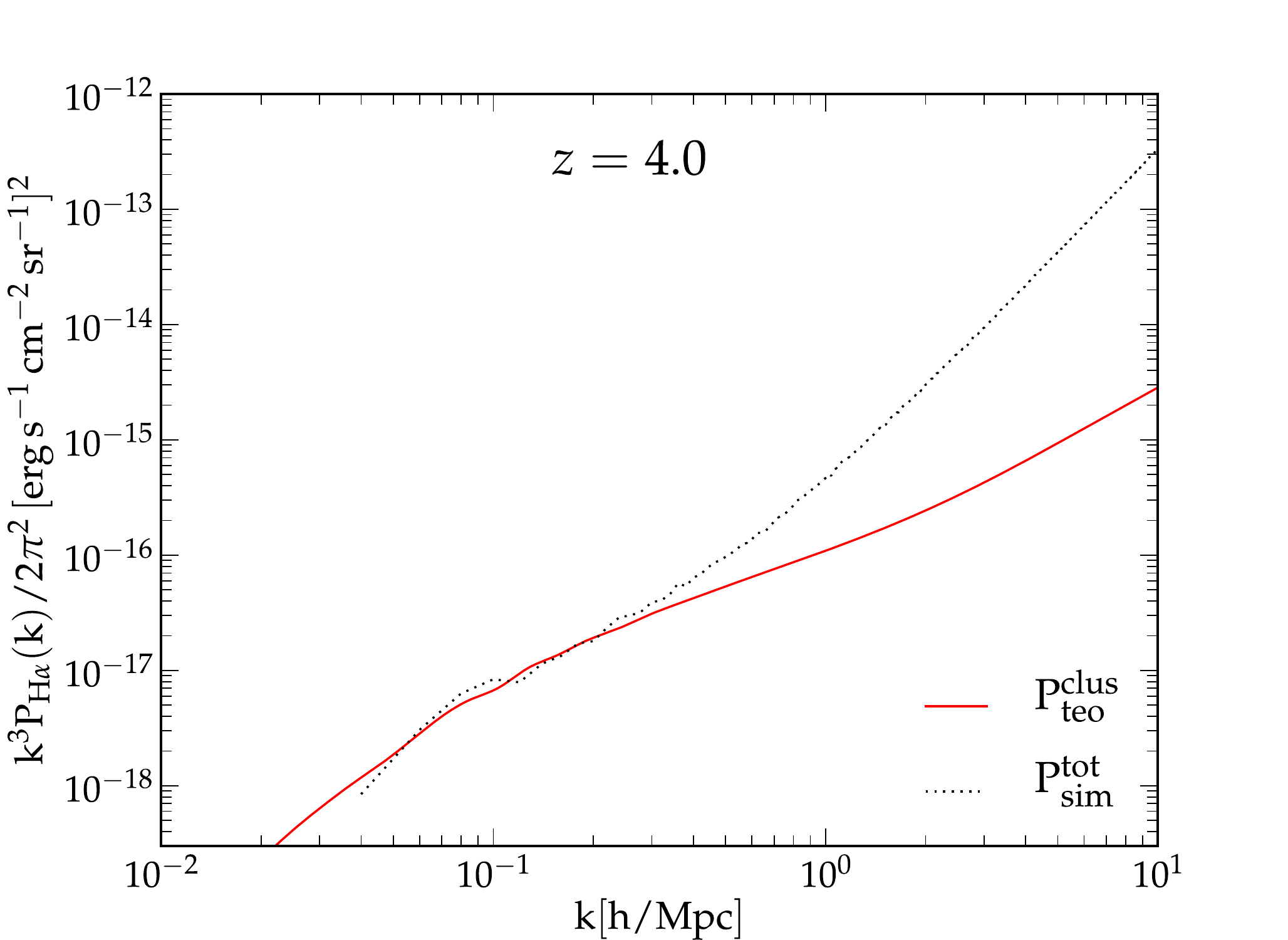}}
}
\caption{Simulated power spectra of \Ha emission from galaxies as it will be observed by the IM surveys (black dashed lines).   
The solid lines show the theoretical clustering power spectrum associated with each survey.}
\label{fig:ps_Ha}
\end{figure*}

Power spectrum analysis is the most common statistic with which intensity maps are studied. 
Figure~\ref{fig:ps_Ha} shows the power spectra of simulated \Ha emission from galaxies at different redshifts. The theoretical \Ha power 
spectrum is also presented in this figure in order to show this line power at large scales. The power spectra in Figure~\ref{fig:ps_Ha} 
correspond to emission from galaxies in DM halos with masses above $6.5\times 10^9\, {\rm M_{\odot}}$. Our theoretical estimates indicate that even at $z\sim 5$, where the emission from low mass halos is more important, their relative contribution to the total \Ha intensity is of the order of 5\%. Therefore, there should be no meaningful loss in power due to the lack of emission from lower mass halos in our simulations.

The IGM \Ha emission is characterized by a small bias (from our simulations we obtain $b_{{\rm H}\alpha}^{\rm IGM}\sim 1$, consistent 
with most of the emission originating in low-density gas). Given the low intensity of the IGM emission, its power spectrum amplitude 
is much smaller than that of galaxy emission. We, therefore, do not show the IGM power spectrum in Figure~\ref{fig:ps_Ha}.

\section{Surveys of \Ha emission} 
\label{sec:Surveys}

Here we make predictions for four instruments: Euclid and WFIRST that will measure \Ha emission from resolved galaxies, and SPHEREx and CDIM that will operate as 
\Ha intensity mapping probes. The properties of the spectroscopic surveys of these instruments are listed in Table \ref{tab:instruments}. 

Euclid is a space mission under development by the European Space Agency that should start observing galaxies in 2020. 
WFIRST is a competing experiment being developed by NASA. It is projected to be launched also in 2020.
In addition to their spectroscopic capabilities, the two satellites will carry out photometric surveys that will cover slightly higher 
frequency/redshift ranges than their
spectroscopic surveys. More importantly, the photometric surveys will allow to probe galaxies much deeper in magnitude. Euclid will reach $\sim$24 mag in $5\sigma$ 
detections of point sources using three filters (Y, J and H) and WFIRST will reach close to 26.5 mag using four filters (Y, J, H and F184).
The frequency resolution of these photometric surveys is enough to probe and 
constrain cosmology using the BAO scale. However, due to the short time scales associated with the astrophysical 
processes at hand, \Ha emission is better constrained using the spectroscopic surveys of these satellites.   

The SPHEREx instrument is meant to be used as an explorer and so it has a broad range of science goals
spanning from cosmic inflation via non-Gaussianity to galaxy evolution and
Galactic ices \citep{2016Dore}. On the other hand, the CDIM instrument is a NASA Probe that focuses on greatly improving our knowledge of galaxy
formation and evolution. This mission's main objective is to probe galaxies and IGM emission during the Epoch of Reionization. The SPHEREx mission
is at an advanced stage of formulation  and was selected for a Medium-Class Explorers Mission concept study by NASA in 2017. It will be a shallow 
all-sky survey but with deep imaging data collected in the ecliptic poles, where the poles are
imaged at every orbital pass. In this study we will only consider the SPHEREx deep survey since its field of view is wide enough to constrain \Ha emission. CDIM remains a concept study and its
exact survey strategy as well as final details are yet to be determined.

Intensity mapping survey data consist of three-dimensional intensity maps, where each observational voxel
is set by a certain angular resolution and covers a wide range of frequency. Given the large volume covered by each observed voxel, it will contain emission from several unresolved sources. Instead of resolving galaxies, IM surveys aim at detecting the overall emission from both bright and faint sources, as 
well as extended emission sources. SPHEREx and CDIM plan to detect \Ha emission in IM mode in order to be 
unconstrained by flux limits and therefore probe the full signal. In these surveys, the recovery of the target signal, i.e., the intensity and spatial fluctuations of \Ha emission, is usually done through power spectrum analysis. However, the same instruments can also operate as galaxy surveyors. As such, they will characterize the galaxies down to the flux limit listed in Table~\ref{tab:instruments}. In this case, due to time limitations, SPHEREx and CDIM will observe over smaller areas.

IM surveys are a better choice for cosmological purposes since they can cover large volumes in a short time. On the other hand, in traditional galaxy survey mode, SPHEREx and CDIM instruments will be able to map bright interloping lines. Furthermore, in the case where the foreground removal strategies in intensity maps are not successful, these instruments will still be able to use the signal from resolved galaxies for astrophysical purposes. 

The large resolving power of galaxy surveys will allow them to provide detailed information on the emission from 
different galaxy populations. The Euclid and WFIRST galaxy surveys will probe the \Ha LF.
On the other hand, SPHEREx and CDIM IM surveys will provide measurements of the integrated \Ha intensity. 
Therefore, IM surveys will probe the entire galaxy population and 
suffer a lot less from selection biases than galaxy surveys.
An additional advantage of IM surveys is that the 
redshift of the source of emission is obtained automatically. This
makes them particularly useful for probing the time evolution of global galaxy properties.
Moreover, given the large frequency range spanned by the SPHEREx and CDIM surveys, they will be able to target several emission lines and to probe 
galaxy emission at higher redshifts than Euclid and WFIRST. 
In particular, the CDIM surveys will cover the high frequencies corresponding 
to Ly$\alpha$ emission from the EoR. CDIM also has the advantage of having a frequency resolution that  makes it
possible to separate between emission in the \Ha line and
in the nearby NII doublet lines.

The potential of each type of instrument to probe galaxies \Ha emission and to constrain different galaxy properties as well as their 
redshift evolution is described in detail in Appendix~\ref{app:constraints}.  
  
\section{Contamination in \Ha observations} 
\label{sec:LineContam}

In both galaxy surveys and IM surveys, observations of \Ha emission will be contaminated by emission from interloping lines. This contamination needs to be 
identified, evaluated and, if necessary and possible, removed. 
 
Euclid and WFIRST low-resolution spectra will be fitted 
with galaxy SED templates, which can, in case the signal to noise is high, be used to identify the observed line. This will, however, not 
always be possible due to the narrow frequency range covered by these instrument's filters. Also, the lack of information on the redshift of the source might result
in line confusion. Moreover, neither of these two surveys has enough frequency resolution to separate the peak of the \Ha line from that of the  NII line doublet. 
Nevertheless, the bright galaxies detected by these surveys are important to constrain the physical properties of the \Ha emitters.  These galaxies will also be
useful to determine the role of the environment in the extinction suffered by \Ha emitters. 
 
Line contamination is also a problem for IM surveys, given that they are intrinsically characterized by detecting emission from all types of 
unresolved sources.
The amount of contribution from line contaminants  depends on the target line and on the observed frequency. Therefore, it has to be evaluated 
according to the frequency range covered by the survey. In the 
case that this contamination is higher or comparable to the signal from the target line, part of it needs to be removed (masked) from the observational maps. 
Generally speaking, the \Ha intensity maps will be contaminated by several strong interloping lines, hence, their contamination needs to be accounted for.
The observational voxel size will determine the percentage of voxels that need to be masked in order to efficiently reduce the contamination in the maps. 
In the case where a high portion of the voxels is masked ($\gtrsim10\%$), the recovery of the target signal might require a correction 
for the loss of flux in the target line. Therefore, 
the appropriate contamination removal strategy for an intensity mapping study is highly dependent on the survey properties.

In Subsection~\ref{sec:contamination1} we model the intensity and dust extinction suffered by each of the line contaminants. For the 
estimation of the observed intensity, we take into account that galaxy surveys will mainly observe bright systems, whereas 
IM surveys are expected to observe the total galaxy population. We continue by estimating the power spectra of line contamination, which is 
relevant for IM surveys (subsection~\ref{sec:contamination2}). We find that the contamination power spectrum is of the same order as that of the signal.
Using simulations we determine the masking fractions required to reduce the 
contamination power spectra to a level well below the predicted \Ha signal. We also present estimates of the masking fractions associated with 
increasing flux cuts (subsection~\ref{sec:contamination3}). 
We assume that the voxels that need to be masked, corresponding to bright foreground emission, will be identified independently by a galaxy survey. 
In the case of the SPHEREx and CDIM missions, the foregrounds survey can be performed by the same instrument operating in a different mode.   

Additionally, intensity maps will suffer from continuum emission originating in the stellar and AGN continuum, as 
well as: free-free, free-bound, 2-photon emission and dust emission in the ISM. At the frequency range relevant 
for \Ha IM, this continuum contamination is dominated by stellar emission, since AGN emission 
only dominates the extragalactic continuum background at higher frequencies/energies \citep{2016Silva}.
Continuum emission is expected to vary more smoothly with frequency compared to the signal from emission lines, which should quickly fluctuate in the frequency direction \citep{2015MNRAS.447..400A}. This smoothness is used in IM studies 
to fit and remove the continuum emission. Therefore, we assume that continuum emission can easily be fitted out of these maps and focus only on line contamination.  The case for galactic contamination is similar, since these foregrounds are fitted in frequency and removed in the same way as extragalactic continuum foregrounds. 

\subsection{Observed intensity of line contaminants}
\label{sec:contamination1}

In \Ha intensity maps, at $z<5.0$, the main contaminants include: the ionized oxygen [OII] 372.7 nm and [OIII] 500.7 nm 
lines, and the hydrogen H$\beta$ 486.1 nm and \lya 121.6 nm lines. We also consider the contamination by 
the [NII] 658.3 nm/654.8 nm and [SII] 671.7 nm/673.1 nm doublet lines. 

Table~\ref{tab:line_cont} presents a list of the contaminating lines and the 
redshifts from which they originate, in comparison to the redshift of the \Ha line. Given 
the very small frequency separation, the NII doublet lines originate from approximately the same redshift as the \Ha line and so they are not included in this table.   
 
In order to model the contamination by each interloping line, we take the published relations between line luminosity
and SFR and then compare/adjust them to the  existing LF constraints, when possible. The intensity of these contaminating lines 
can be calculated by integrating over the SFR or the line LF, in the same way as was done for the \Ha line, using Equation~\ref{eq:I_LF}.

In most cases, the intensity of the interloping lines are only constrained at low redshifts and the extrapolation of their LFs to higher redshifts are uncertain. 
The several factors involved in the line LF redshift evolution include: the expected decrease in galaxy metallicity, 
the lower ionization state of the ISM gas, the possibility of the existence of low-density canals which would 
affect the galaxy extinction rates independent of the galaxy dust content, and others.

We now explore the contamination from each line in further detail.\\

\begin{table*}
\centering            
\caption{Contaminant background lines in Halpha intensity maps}            
\begin{tabular}{l  c c c c c c c}        
\hline\hline                 
Line  & $\lambda_0\, {\rm [nm]}$    & $z$($\lambda=656.28\, {\rm nm}$) & $z$($\lambda=721.91\, {\rm nm}$) & $z$($\lambda=1181.30\, {\rm nm}$) & $z$($\lambda=1968.84\, {\rm nm}$)  & $z$($\lambda=2625.12\, {\rm nm}$)  & $z$($\lambda=3937.68\, {\rm nm}$)\\    
\hline                 
   ${\rm H}_{\alpha}$ & 656.28  & 0.00  & 0.20  & 0.80  & 2.00   & 3.00   &  5.00 \\
   SII                & 671.7   & ...   & 0.07  & 0.76  & 1.93   & 2.91   &  4.86 \\
   OIII               & 500.7   & 0.31  & 0.57  & 1.36  & 2.93   & 4.24   &  6.86 \\
   ${\rm H}_{\beta}$  & 486.1   & 0.35  & 0.62  & 1.43  & 3.05   & 4.40   &  7.10 \\
   OII                & 372.7   & 0.76  & 1.13  & 2.17  & 4.28   & 6.04   &  9.57 \\
   Lya                & 121.6   & 4.40  & 5.48  & 8.71  & 15.19  & 20.59  &  31.38 \\
  \hline                                  
\end{tabular}
\label{tab:line_cont}     
\end{table*}

\textbf{NII contamination:} The relative contribution from the NII doublet to the \Ha plus 
NII lines scales with the equivalent width (EW) of this peak and is usually in the range 10\% to 50\% \citep{2009Sobral,2012Sobral}. 
As a reference, the average value of the NII line contribution at $z=1.47\pm0.02$ and for \Ha fluxes above 
$7\times 10^{-17}\, {\rm erg\, s^{-1}\, cm^{-2}}$ is $\sim$ 22\% of the sum of the \Ha and the NII line intensities \citep{2012Sobral}.

For galaxy surveys, we assume that the NII line contamination can be estimated from the \Ha EW following the 
relation observed in SDSS galaxies and parameterized by \citet{2012Sobral} as:
\ba
{\rm log\left([NII]/H\alpha]\right)}&=&-0.924+4.802E-8.892E^2\nonumber \\
&+&6.701E^3-2.27E^4+0.279E^5,
\ea
where  $E{\rm = log\left[ EW_0 (N_{\rm II}+H\alpha) \right]}$. 

However, IM surveys do not resolve individual galaxies. Therefore, this parametrization is impossible to apply. Given that,  in intensity maps, each voxel contains
emission from several galaxies, our default assumption is that the NII line contributes $22\%$ of the sum of the \Ha and NII line intensities. 
A similar NII line contribution is 
predicted by Galaxy Evolutionary Synthesis (GALEV) models for galaxies with solar metallicities \citep{2003Anders}. For sub-solar metallicities 
the NII contribution to the NII plus \Ha line intensities should be smaller.
One of the main advantages of the CDIM surveys is that they will have sufficient frequency resolution 
to distinguish between the \Ha line emission and the NII line doublet emission. 

From the surveys we consider in this work, 
only CDIM will have enough frequency resolution to do this. It should be noted that the evolution of the ${\rm NII/H\alpha}$ at $z\gtrsim1.47$ 
is uncertain, due to lack of observations. Therefore, the assumption that this ratio is fixed will add a large systematic error to the estimation of the \Ha signal.
For further discussion on this topic see Appendix~\ref{subsec:Z_ion_param}. \\

\textbf{SII contamination:} Most surveys also do not have good enough frequency resolution to separate the peak of the \Ha line 
from that of the SII line doublet. Therefore, one needs to estimate the average contribution of this doublet line, 
using the same method we applied to the NII line.

The strength of the SII doublet has been found to vary between $\sim10-60\%$ of the \Ha emission in SDSS star-forming galaxies 
\citep{2006Kewley}. However, in most cases the contribution of the SII lines to the (\Ha + NII + SII) peak is $\sim$12\%. This 
value is taken from \citet{2016Marmol}, who evaluated it using \Ha selected galaxies in the redshift range 
$1.23\lesssim z\lesssim 1.49$ and with an average equivalent width of $175\pm14\AA$ .

The contribution from these doublet lines should be estimated directly from the survey data. Whenever that is not possible, we assume
that the SII doublet line intensity is of the order of 12\% of the \Ha + NII+ SII line intensity. We note that GALEV models for galaxies with solar 
metallicities predict that the SII line intensity corresponds to 9\% of the total \Ha + NII + SII line flux \citep{2003Anders}.\\

\textbf{H$\beta$ contamination:} We estimate the H$_{\beta}$ intrinsic luminosity from Eq.~\ref{eq:Lum_Ha} and the recombination emission line ratios, which yields,
\be
L^{\rm int}_{\rm H\beta}\, ({\rm erg\, s^{-1} })\, =\, 4.43\times 10^{40} SFR\, ({\rm M_{\odot}\, yr^{-1}}).
\label{eq:Lum_Hb}
\ee

Based on the \citet{2000Calzetti} extinction curve and the dust attenuation by OB galaxies in the COSMOS survey up to $z\, \sim\, 6.5$ \citep{2015Scoville}, 
the extinction suffered by the \Hb line is of the order of $A_{\rm H\beta}=1.35\times A_{\rm H\alpha}$. We use this to estimate the average \Hb intensity.   

\begin{figure}
\begin{centering}  
\includegraphics[angle=0,width=0.5\textwidth]{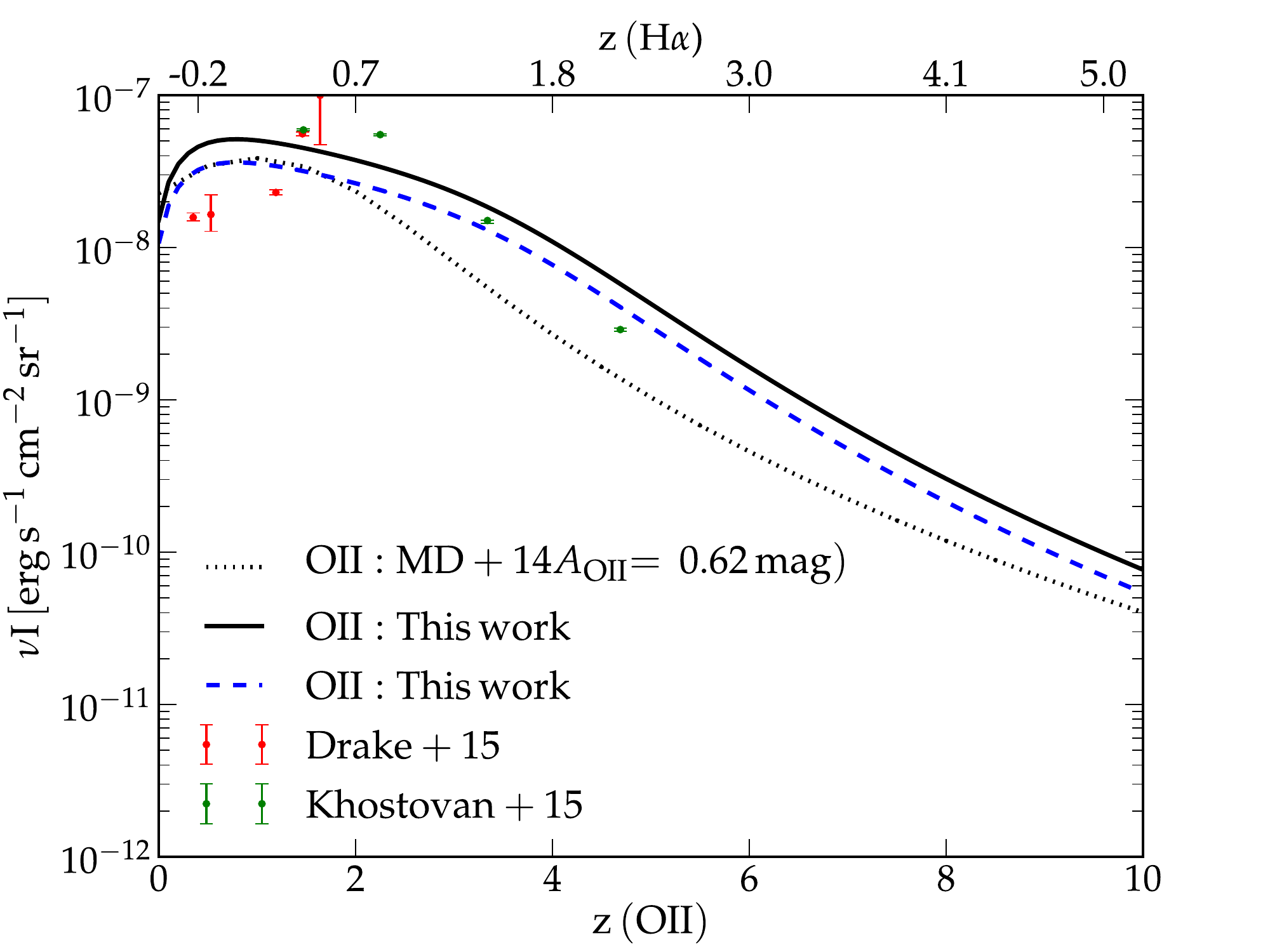}
\caption{Intensity of observed OII line emission as a function of redshift. The lines show the OII intensity derived from our SFRD model 
(solid line and dashed line) and from 
the MD+14 SFRD model (dotted line). The solid and the dotted black lines were corrected for a 
dust extinction of $A_{\rm OII}\,=\, 0.62\, {\rm mag}$, while the blue dashed line was corrected for a dust extinction of $A_{\rm OII}\, =\, 1.0\, {\rm mag}$. 
The dots correspond to observational points from UKIDSS \citep{2013Drake} and HiZELS \citep{2015Khostovan}.}
\label{fig:I_OII}
\end{centering}
\end{figure}

\begin{figure}
\begin{centering}  
\includegraphics[angle=0,width=0.5\textwidth]{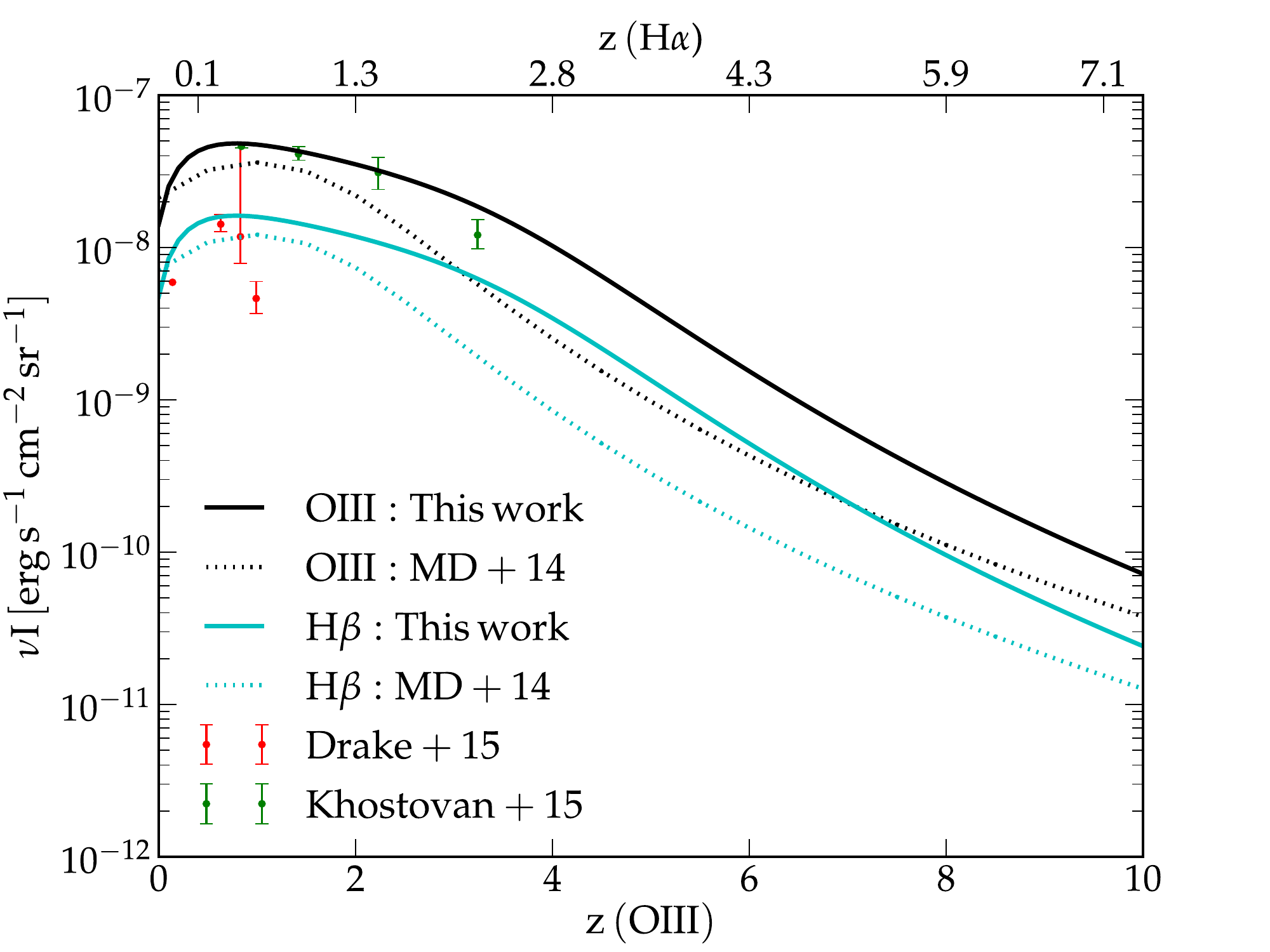}
\caption{Intensity of observed OIII line emission as a function of redshift. The black solid and the black dotted lines 
show, respectively, the OIII intensity derived from our SFRD model and from 
the MD+14 SFRD model. To the intrinsic line intensities, we applied a dust extinction of $A_{\rm OIII}\, =\, 1.35\, {\rm mag}$. 
The dots correspond to observational points from 
the UKIDSS Ultra Deep Survey Field \citep{2013Drake} and from HiZELS \citep{2015Khostovan}.  Also shown in the 
cyan lines is the H$\beta$ intensity obtained using Equation~\ref{eq:Lum_Hb}, to which we applied  
a dust extinction of $A_{\rm H\beta}\, =\, 1.35\, {\rm mag}$. The cyan solid line assumes our SDRD model while the cyan dotted lines uses the MD+14 SFRD model.}
\label{fig:I_OIII}
\end{centering}
\end{figure}

\textbf{OII contamination:} The OII line luminosity can be estimated from the galaxy SFR using existing fits to the 
observational data. As a reference, we use the relation from 
\citet{1998KennicuttAraa} based on a ratio of 0.57 between the OII and \Ha fluxes observed in local galaxies:
\be
L_{\rm OII}\,  ({\rm erg\, s^{-1}})\, =\, (7.18\pm 2.2) \times 10^{40}\, SFR\, ({\rm M_{\odot}\, yr^{-1}}).
\label{eq:LOII_SFR}
\ee
In Figure~\ref{fig:I_OII}, we compare the OII intensities predicted by SFRD based models with observational 
constraints from the UKIDSS \citep{2013Drake} and HiZELS \citep{2015Khostovan} surveys. At low $z$, the theoretical 
curves are systematically above the predictions from \cite{2013Drake}. However, at $z\sim1.5-2$ our predictions fit quite well. As indicated 
by the more recent measurements from \cite{2017Matthee}, this reflects a lower OII$/$\Ha ratio at low $z$. The evolution of this ratio can be due 
to an increasing dust extinction at high $z$. It can also simply be due to technical issues with observations in the local Universe given that the 
surveys only measure the central kpc region of the galaxies \cite{2017Matthee}. 

The high end of the OII LF is observationally probed up to a redshift of $z\, \sim\,5$. However, this line intensity is only well 
constrained up to $z\, \sim\, 1.5$ \citep{2015Khostovan}. Recent observational 
constraints, reaching $z\sim4.5$, predict this line EW to increase up to 
$z\sim3.5$ and then to decline as a result of a higher ionization state of the ISM \citep{2016Khostovan}. 

The OII intensity in this figure assumes a low 
dust extinction of $A_{\rm OII}\, =\, 0.62$ mag \citep{2013Hayashi.Sobral}. This is the average dust extinction 
of the observed OII emitting galaxies. However, the relevant extinction for IM studies should be that of the full 
galaxy population, which should be closer to $A_{\rm OII}\, =\, 1$ mag, see discussion in \cite{2013Hayashi.Sobral}.\\

\textbf{OIII contamination:} Similarly, the OIII line luminosity can be estimated from the galaxy SFR, using existing fits to observational data. We take the 
relation from \citet{2007Ly.Malkan}, based on observations in the $z\sim 0.07-1.47$ redshift range and given by:
\be
L_{\rm OIII}\, ({\rm erg\, s^{-1}})\, =\, (1.32\pm 2.7) \times 10^{41}\, SFR\, ({\rm M_{\odot}\, yr^{-1}}).
\label{eq:LOIII_SFR}
\ee
In Figure~\ref{fig:I_OIII}, we compare the OIII intensities predicted by SFRD based models with observational 
constraints from the UKIDSS \citep{2013Drake} and HiZELS \citep{2015Khostovan} surveys. To the OIII and the H$\beta$ line intensities, we applied a dust extinction 
of $A_{\rm OIII}=A_{\rm H\beta}=1.35\, {\rm mag}$, which corresponds to $A_{\rm H\alpha}\, =\,1\, {\rm mag}$ \citep{2015Khostovan}. 
As in the case of the OII line, our theoretical OIII line curves fit better with observational LFs at $z\gtrsim1.5$. We note, however, 
that the observational line ratios
OIII/OII and OIII/\Hb are larger in high $z$ galaxies \citep{2017Castellano}. 
Also, the EW of the OIII line should increase towards galaxies with small 
stellar masses. Therefore, these high EW are increasingly important towards high $z$, where galaxies are on average 
smaller \citep{2016Khostovan}. This indicates that the  ISM in high $z$ 
galaxies is characterized by a higher ionization parameter than that of low $z$ galaxies.
  
With the few currently available observations it is still unclear whether these differences are due to redshift evolution
of the ratios between lines, dust extinction or technical differences in observations, such as aperture effects \citep{2017Matthee}. Nonetheless, in the redshift 
range important for this study, predictions based on observations fit well with those from our SFRD based model. Therefore, we estimate the contamination 
by OIII line emitters using the latter model.  \\

\textbf{\lya contamination:} \lya from high redshift galaxies, $z > 4.4$, contaminates  \Ha line observations in the redshift range $z\sim 0\, -\, 5$. 

We take the \lya LFs at $z\sim5.7$ and $z\sim6.6$, from \citet{2016Santos.Sobral}, and integrate them down to 
$L_{\rm Ly\alpha}=10^{38}\, {\rm erg\, s^{-1}}$. For the different possible values of the low luminosity slope of the \lya LF, the  
observed line intensity is between $(0.13-3.9)\times10^{-8}\, {\rm erg\, s^{-1}\, sr^{-1}\, cm^{-3}}$ at $ z=5.7$
and between $(0.036-1.1)\times10^{-8}\, {\rm erg\, s^{-1}\, sr^{-1}\, cm^{-3}}$ at $ z=6.6$. This high uncertainty 
is due to, at these redshifts, the \lya LF only being observationally constrained down to a luminosity of $\sim10^{42.5}\, {\rm erg\, s^{-1}}$.

The \lya intensity can also be estimated from the SFR as
\be
L_{\rm Ly\alpha}\,  ({\rm erg\, s^{-1}})\, =\, 1.1 \times 10^{42} \,SFR\, ({\rm M_{\odot}\, yr^{-1}}).
\label{eq:Lya_SFR}
\ee 
This observational relation was obtained using galaxies in the local Universe \citep{1998KennicuttAraa}. 

Using our SFR model the \lya line intrinsic intensity is then $6\times10^{-8}\, {\rm erg\, s^{-1}\, sr^{-1}\, cm^{-3}}$ 
at $z = 5.7$ and $2.6\times10^{-8}\, {\rm erg\, s^{-1}\, sr^{-1}\, cm^{-3}}$ at $z = 6.6$.

The observed line intensity was obtained by correcting the \lya line intensity for the average of the \lya photon escape fraction ($f_{\rm esc}^{\rm Ly\alpha}$). 
Due to the lack of good observational, for the \lya dust extinction in galaxies, we take $f_{\rm esc}^{\rm Ly\alpha}$ from the high-resolution 
simulations of \citet{2014Yajima}. As a reference this study predicts the average value of this escape fraction to be $f_{\rm esc}^{\rm Ly\alpha}(z=5.5)\sim0.5$.

The \lya intensity estimated from LFs is smaller than that estimated using our SFR model. LFs are obtained using galaxy survey data which are 
unable to detect scattered \lya photons around the galaxy. These photons will, however, be detectable by IM surveys. Therefore, we estimate 
this line contamination in intensity maps using our SFR estimate.    

There will be additional contamination in the \Ha intensity 
maps from \lya emission  originating from recombinations, collisional 
excitations and scattering of Lyman-n photons in the IGM \citep{2013Silva, 2016Comaschi}.  
Even for the higher estimates for this scattered emission obtained by \cite{2016Comaschi}, its power spectrum will 
be considerably below the signal from galaxies.
Moreover, the bias associated with the IGM emission is very close to unity, so in power spectrum analysis, these IGM 
contributions are not important. For simplicity, we, 
therefore, ignore these contributions to the \lya power spectrum. 

\begin{figure*}
\vspace{-4pt}
\centerline{
\resizebox{!}{!}{\includegraphics[angle=0,width=0.50\textwidth]{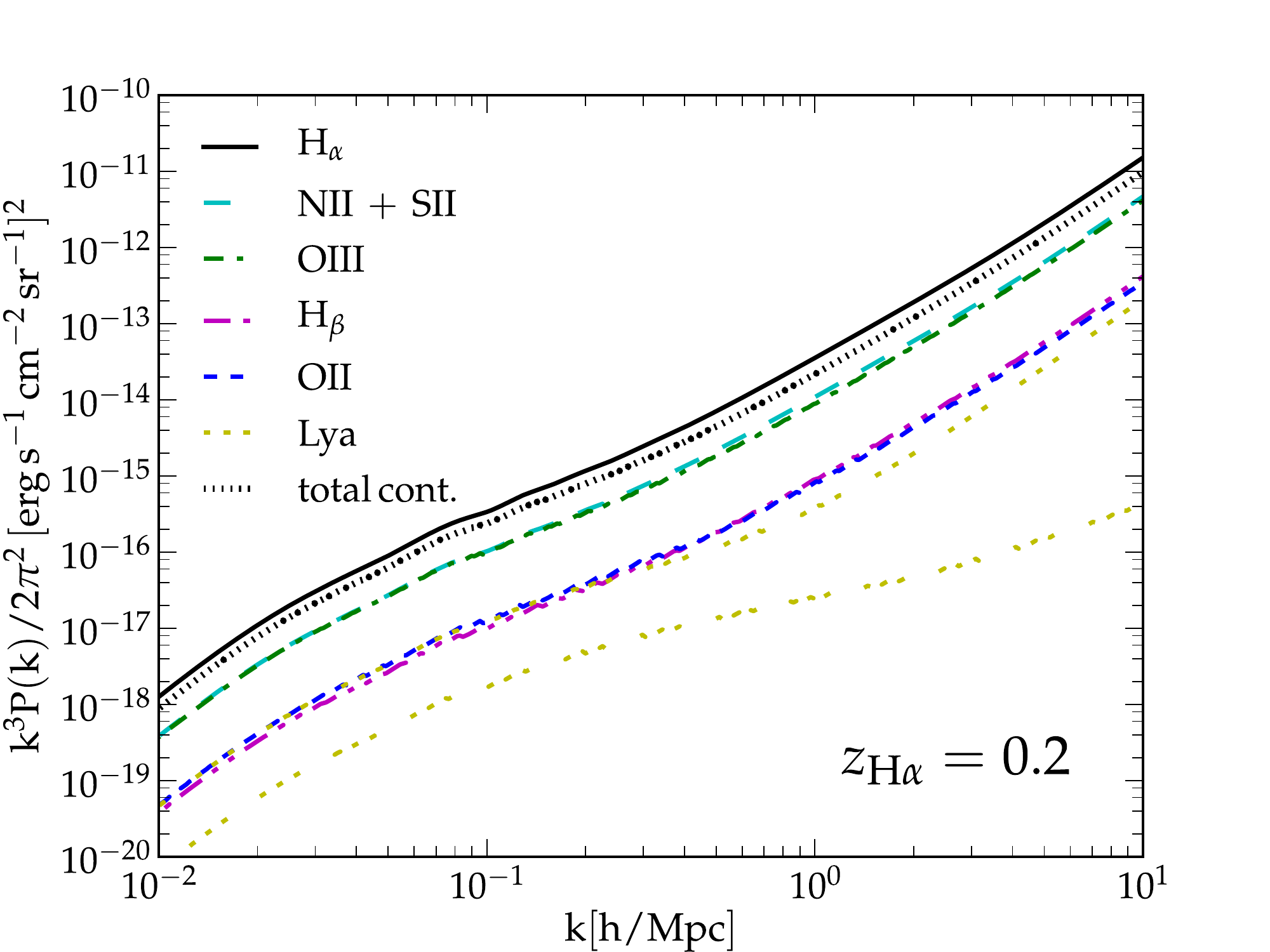}}
\resizebox{!}{!}{\includegraphics[angle=0,width=0.50\textwidth]{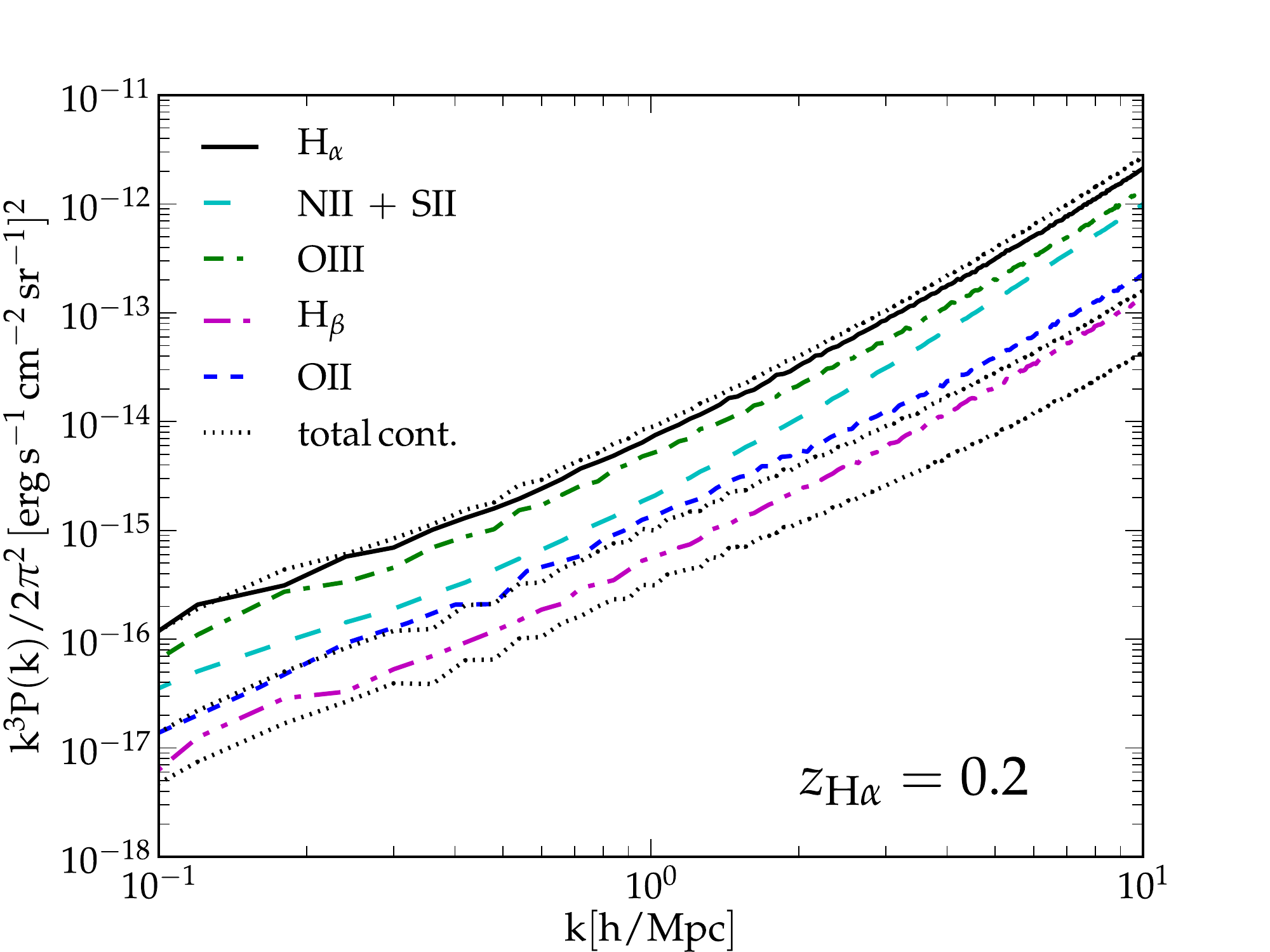}}
}
\vspace{-1pt}
\centerline{
\resizebox{!}{!}{\includegraphics[angle=0,width=0.50\textwidth]{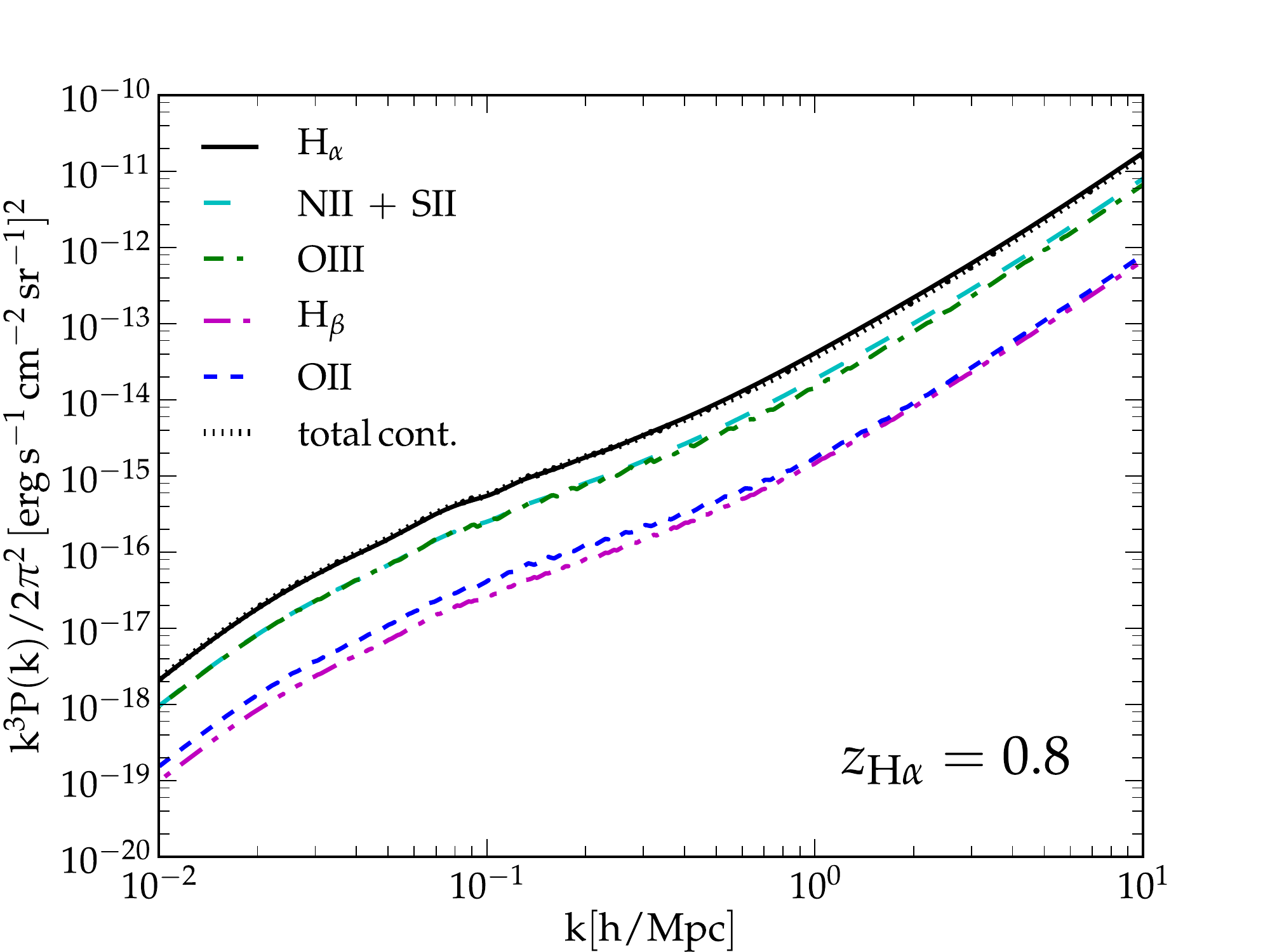}}
\resizebox{!}{!}{\includegraphics[angle=0,width=0.50\textwidth]{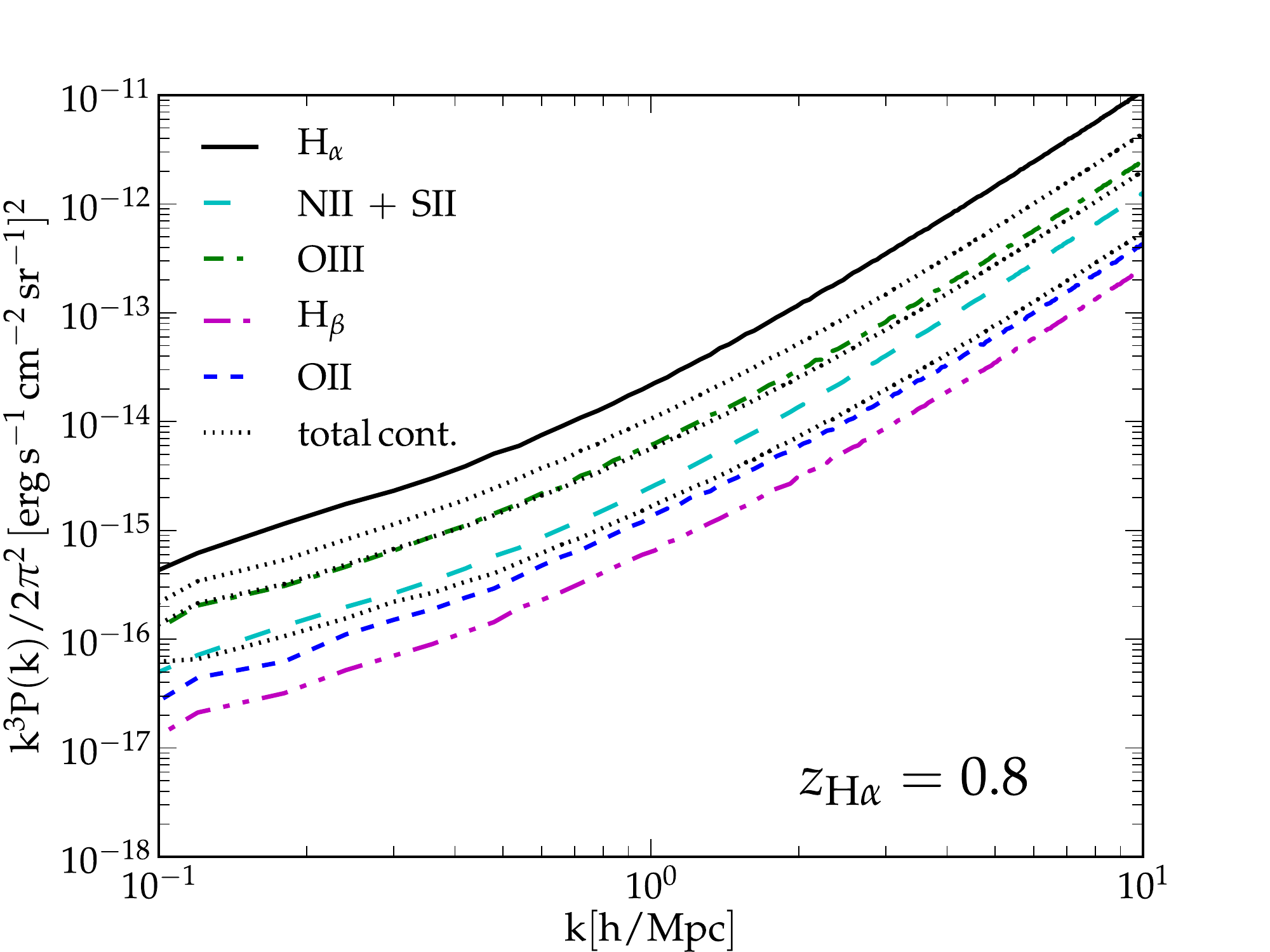}}
}
\vspace{-1pt}
\centerline{
\resizebox{!}{!}{\includegraphics[angle=0,width=0.50\textwidth]{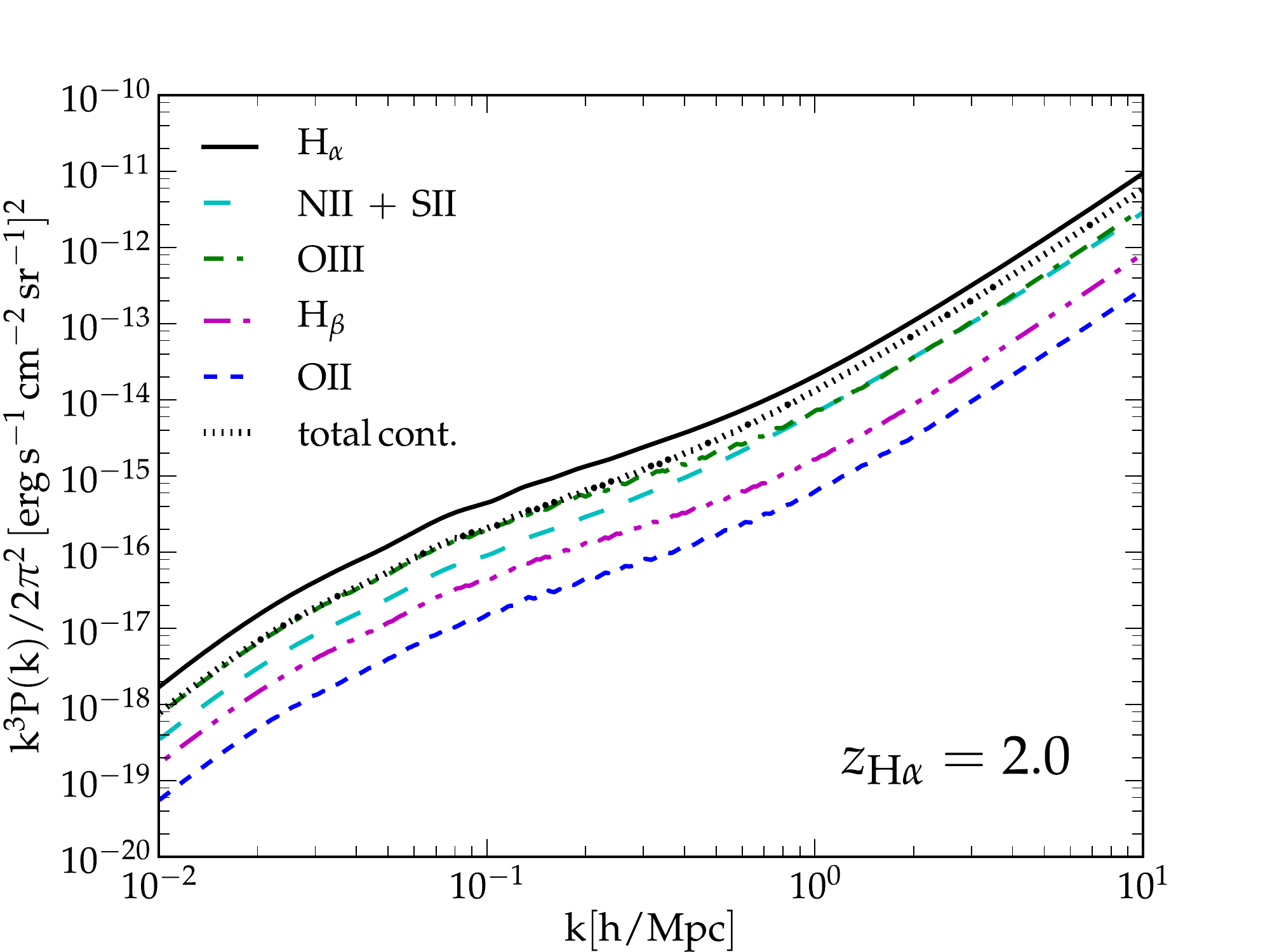}}
\resizebox{!}{!}{\includegraphics[angle=0,width=0.50\textwidth]{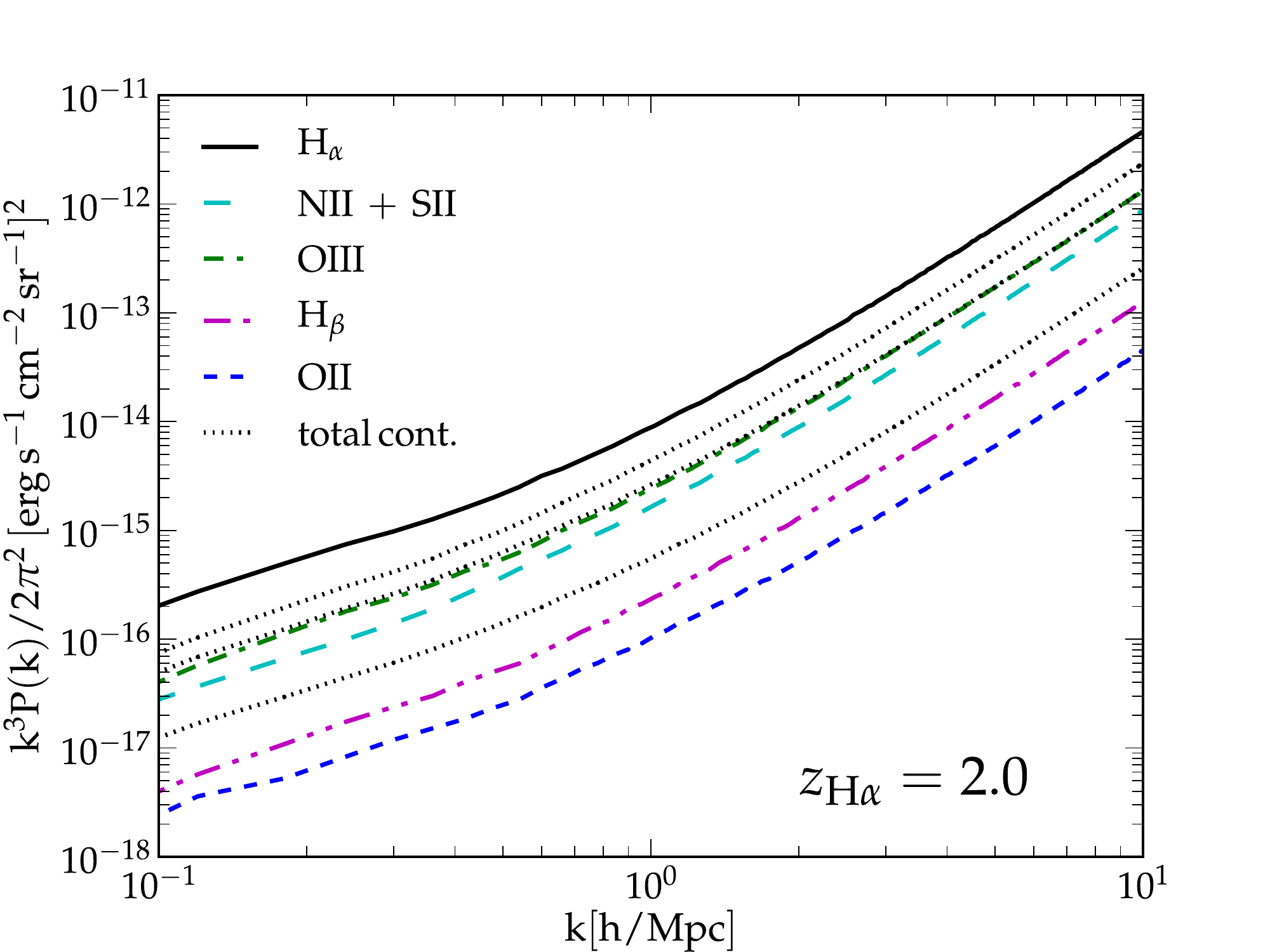}}
}
\caption{
Left panels: Theoretical power spectra of contamination by background lines in \Ha intensity maps at $z_{{\rm H}\alpha}\, =\, 0.2, 0.8, 2.0$ 
(from top to bottom). The power spectra from the several interloping lines were scaled to the redshift of the \Ha line. The upper three 
dotted yellow line corresponds to \lya emission from galaxies.
The IGM \lya emission power spectrum (bottom three dotted yellow line) was calculated using: an intensity with double the value of the intensity 
from galaxies, a bias with the underlying density of one and no shot noise.
Right panels: 
Simulated power spectra of contamination by background lines in \Ha intensity maps at $z_{{\rm H}\alpha}\, =\, 0.2, 0.8, 2.0$ 
(from top to bottom). These 
plots were obtained from our simulation and assume the cell resolution of the CDIM survey. The result would be very similar for a survey with the 
spatial and frequency resolution of SPHEREx. The top dotted line denotes 
the total contamination power  spectrum (by the OIII, OII, and H$_\beta$ lines), while the remaining dotted lines denote the contamination power 
spectra, after masking cells with fluxes in one of the foreground lines above a given threshold. For $z=0.2$ and $z=0.8$ the flux thresholds 
for masked cells are $1.2\times10^{-16}\, {\rm erg\, s^{-1}\, cm^{-2}}$ (middle dotted line) and $5.0\times10^{-17}\, {\rm erg\, s^{-1}\, cm^{-2}}$ (bottom dotted line). 
For $z=2$ the flux thresholds for masked cells are $5.0\times10^{-17}\, {\rm erg\, s^{-1}\, cm^{-2}}$ (middle dotted line) and 
$1.0\times10^{-17}\, {\rm erg\, s^{-1}\, cm^{-2}}$ (bottom dotted line).}
\label{fig:ps_lines}
\end{figure*}

\subsection{Power spectra of line contamination}
\label{sec:contamination2}

Intensity maps can be analyzed in several ways. However, given the low signal-to-noise ratio in the maps expected in many experiments, the signal 
will most likely be detected statistically. The most obvious statistic to use is the intensity power spectrum \cite[see e.g.,][]{2012Pritchard}. 
 
The target line and the interloping lines are emitted from different redshifts and so their emission originates in different volumes.
One needs to account for a volume conversion factor when estimating the contamination power spectra. 
We estimate the contamination by background lines in the observed \Ha power spectrum as a function of the perpendicular and parallel components of the wavevector k,
using the \cite{2014GongLya} formula, given by: 
\ba 
P_{\rm obs}(k_{\perp},k_{\parallel})&=& \left[ P^{\rm clus}_{\rm line}(z_{\rm f},k_{\rm f}) + P_{\rm line}^{\rm shot}(z_{\rm f},k_{\rm f}) \right] \nonumber \\
 &\times&  \left(\left[ \frac{\chi(z_{\rm s})}{\chi(z_{\rm f})}\right]^2 \left[\frac{y(z_{\rm s})}{y(z_{\rm f})}\right] \right),
 \label{eq:ContPSshift}
\ea
where the clustering power spectrum is,
\ba
P^{\rm clus}_{\rm line}(z_{\rm f},k_{\rm f})&=& \bar{I}^2_{\rm f}(z_{\rm f}) b^2_{\rm f}(z_{\rm f}) P_{\delta\delta}(z_{\rm f},k_{\rm f}).
\ea
The indexes $s$ and $f$ indicate the source, i.e.,
${\rm H}\alpha$, or the foreground/background line redshifts, respectively. The parameter $r$ corresponds to the comoving distance, while
$|\vec{k_{\rm f}}|\, =\, \left[ (r_{\rm s}/r_{\rm f})^2 k^2_{\perp}\, +\, (y_{\rm s}/y_{\rm f})^2 k^2_{\parallel} \right]^{1/2}$ is the three-dimensional k vector at
the redshift of the foreground/background line. $P_{\delta\delta}$ is the matter power spectrum and $b_{\rm line}$ is 
the bias between the interloping line luminosity and the dark matter fluctuations.
Finally, the shot noise power spectrum due to the discrete nature of galaxies is:
\be
P^{\rm shot}_{\rm line}(z)=\int^{M_{\rm max}}_{M_{\rm min}} dM \frac{dn}{dM} \left[ \frac{L(M,z)}{4 \pi D^2_{\rm L}}y(z)D^2_{\rm A} \right]^2.
\ee
Notice that the contaminants power spectrum shown in Eq.~\ref{eq:ContPSshift} experiences a scale-dependent shift, as well as, amplitude modification. 
This change, due to contamination, will increase/decrease the measured power spectrum. In the scales measured by the IM surveys, 
the contamination by background lines will be attenuated relative to the \Ha power spectrum, whereas the foreground interloping 
lines will have the opposite effect. Therefore, most studies ignore the influence of background lines \cite[see e.g.,][]{2014GongLya,2017Fonseca}. However, we show here, after a careful estimation, that some of the background lines cannot be ignored.

Left panels in Figure~\ref{fig:ps_lines} show the power spectrum of the most important background lines that can be confused with the \Ha emission. 
The amplitude of the lines shown in this plot were calculated assuming a survey capable of detecting the full line emission. This is a reasonable 
assumption for IM surveys. The matter 
power spectrum was theoretically estimated using the publicly available code CAMB \citep{2000Lewis}. 

Figure~\ref{fig:ps_lines} clearly shows that the contamination power spectrum from background lines is of the order of the H$\alpha$ power spectrum. 
Therefore, it is clear that this contamination is not negligible and, hence, can not be ignored. Namely, some of this background 
contamination needs to be removed from observational maps in order to accurately recover the signal from \Ha emission. 
We note that the amplitude of the \lya line contamination power spectra are very small and can always be ignored. We show its 
values only in the top left panel of Figure~\ref{fig:ps_lines}, whereas, in the lower panels it is ignored because it is even smaller. 

\subsection{Contaminants masking fractions derived from simulations}
\label{sec:contamination3}
\begin{figure*}
\vspace{-10pt}
\centerline{
\hspace{2pt}
\resizebox{!}{!}{\includegraphics[angle=0,width=0.5\textwidth]{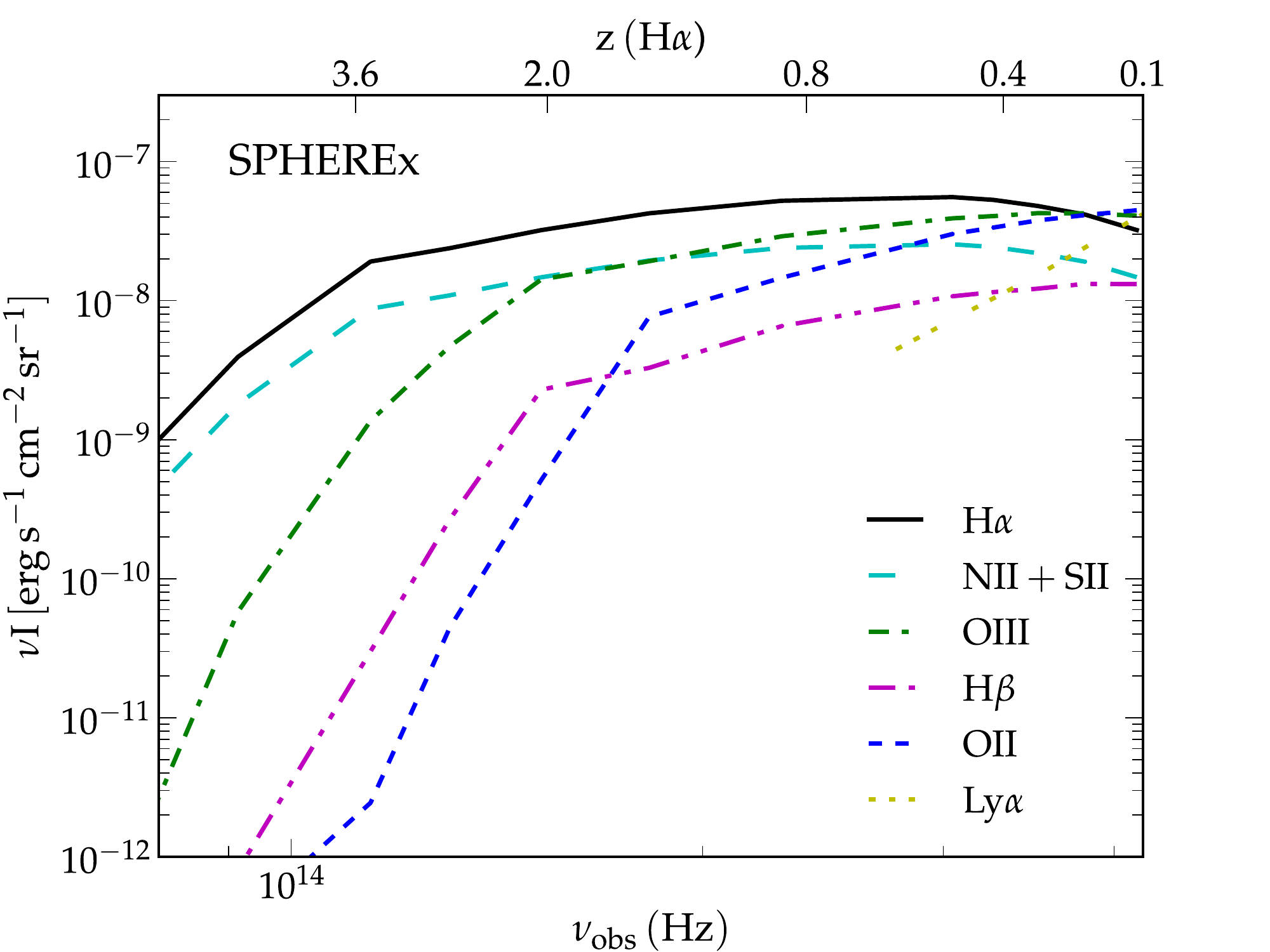}}
\hspace{-12pt}
\resizebox{!}{!}{\includegraphics[angle=0,width=0.5\textwidth]{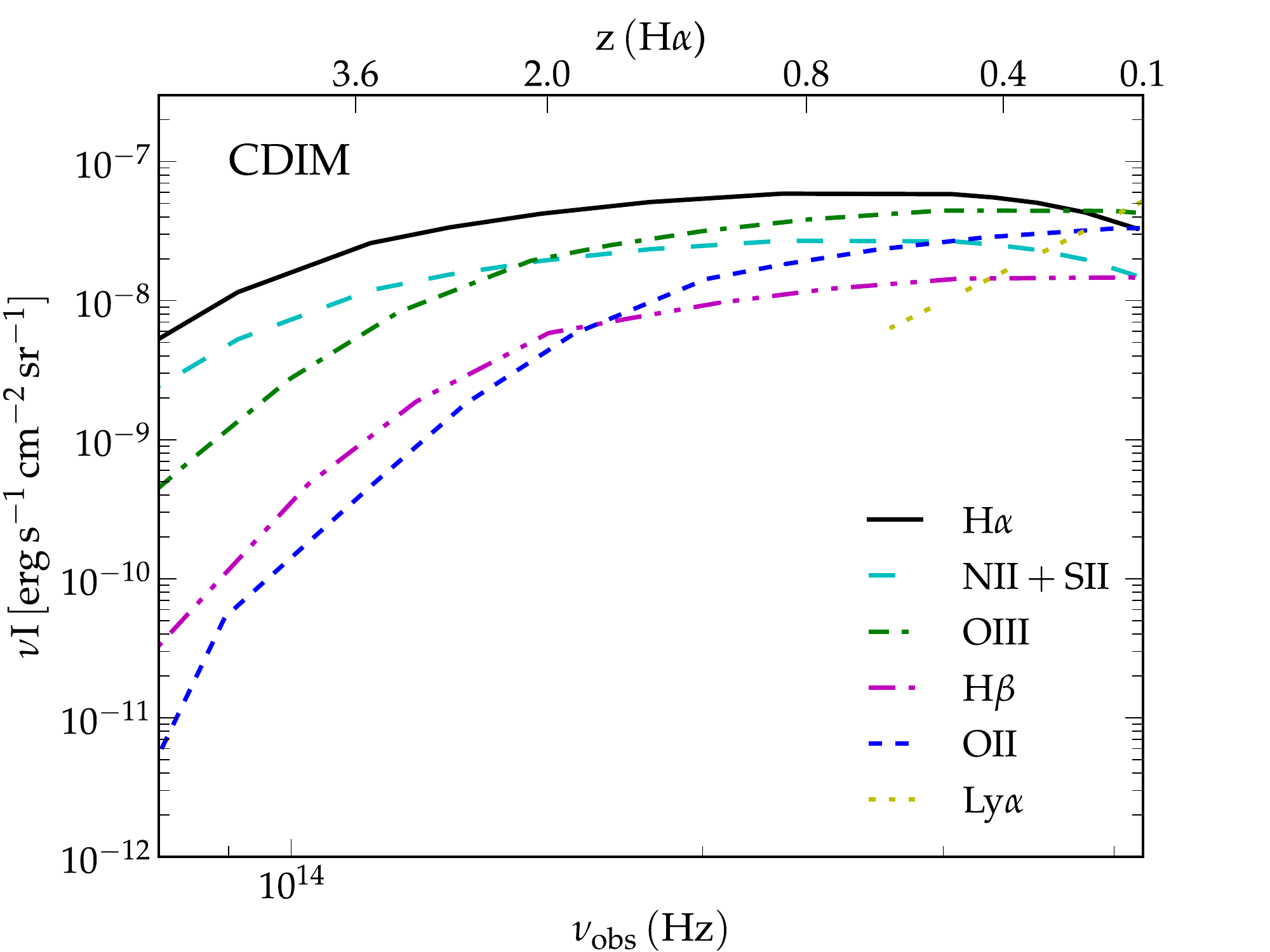}}
}
\caption{Intensity of several emission lines in the $( 0.8- 4.3)\times10^{14}\, {\rm Hz}$ frequency range (corresponding to a 
wavelength range of ${\rm 3.75-0.70\, \mu m}$), as it will be observed by the SPHEREx and CDIM surveys  in galaxy survey mode. These plots were obtained from 
our simulation and assume both the cell resolution and the flux sensitivity of these surveys. Note that for the CDIM survey we assumed a 
flux sensitivity of $10^{-18}\, {\rm erg\, s^{-1}\, cm^{-2}}$. Also, for the NII and SII doublet lines we simply assumed that, 
following the discussion in Section~\ref{sec:contamination1}, they would have 0.282 and 0.176 of the intensity of the \Ha line, respectively. 
The intensity of \lya emission is only shown up to $z\sim8$, given that the intensity of this line at higher $z$ is highly dependent on the 
assumed hydrogen reionization history.}
\label{fig:I_foreg_nu}
\end{figure*} 

We use the simulations described in Section~\ref{sec:Simulations} to estimate the flux cuts that we need to mask 
contaminant line emission, in order to efficiently reduce their power spectra. With the same simulations we 
calculate the percentage of observational voxels that need to be masked to achieve these flux cuts. 

The right panel of
Figure~\ref{fig:ps_lines}, shows the residual line contamination after masking the brightest background emitters from the observational maps. Note that these power spectra
have a bit more power at small scales than the theoretical ones
given that our simulation does not have halos below 
$M_{\rm min}= 6.5\times 10^{9}\, {\rm M}_{\odot}$ and so we are overestimating the shot noise. 
The contamination in \Ha intensity maps is
considerably reduced at $z\sim0.2$ by masking voxels containing the signal from background emitters with fluxes above 
$1.2\times 10^{-16}\, {\rm erg\, s^{-1}\, cm^{-2}}$. This would decrease the contamination power spectrum to about $5-25\%$ of its 
initial amplitude. Moreover, it would only require masking less than $1\%$ percent of the observational voxels, both for SPHEREx and for CDIM. 

At $z\sim0.8$, masking voxels contaminated by line fluxes above $1.2\times10^{-16}\, {\rm erg\, s^{-1}\, cm^{-2}}$ would only slightly 
decrease the signal from background lines. This would leave the line contamination in the observed maps at the level of $10-20\%$ of 
the total observed power spectrum. A flux cut of $5.0\times10^{-17}\, {\rm erg\, s^{-1}\, cm^{-2}}$ would be much more successful at 
reducing this contamination. However, it is very observationally challenging to individually detect all the galaxies responsible for this 
emission down to this low flux level. 

At $z\sim2.0$ and at a scale of $k\sim 0.1\, {\rm h^{-1}\, Mpc}$, the \Ha signal will be approximately $5$ times 
higher than the background lines signal. Meaningfully decreasing this contamination (to about $50\%$ of its initial value) at small scales would also 
require the masking of contaminant lines with fluxes down to $\sim 5.0\times10^{-17}\, {\rm erg\, s^{-1}\, cm^{-2}}$ (see Fig.~\ref{fig:ps_lines}). 

The percentage of voxels lost, assuming these flux cuts and the voxel size of the CDIM and SPHEREx surveys is always at the $0.1-1\%$ level. 
OIII emitters dominate the contamination in these maps. Therefore, detecting and masking only OIII line contaminants would be observationally easier and would still result in a meaningful reduction of the contamination in \Ha intensity maps. 

The quoted masking fractions will only decrease the \Ha 
intensity by less than $1\%$. All these results were confirmed using our simulations. 

An alternative masking procedure is the so called blind masking in which the brightest pixels are masked assuming they belong to foreground galaxies. However, this type of masking will not work in the presence of background line contaminants, since by doing so, a large fraction of the target line would also be inevitably masked. For the previously quoted flux cuts, 
the percentage of \Ha signal that would be erased is of the order of $40-85\%$.

As a side note, contamination by interloping lines causes anisotropies in the angular power spectrum  because the different lines originate at different redshifts. 
The angular power spectrum can, therefore, be used to test if the masking procedure was successful.

\section{Alternative foreground removal/avoiding methods}
\label{sec:contamination4}

In the case where the foreground removal strategies in intensity maps are not successful, CDIM and SPHEREx will still be able to use their deep surveys (in which they function as traditional galaxy surveys and resolve galaxies) for astrophysical purposes. 

In Figure~\ref{fig:I_foreg_nu}, we show estimates for CDIM and SPHEREx constraints on bright line intensities, in order to predict how useful they will be in tracing global astrophysical quantities when operating in galaxy survey mode.
Note that with the assumed flux limit for CDIM, the observed line intensity is very close to its total value.
Figure~\ref{fig:I_foreg_nu} indicates that at $z\sim 0-5$, the intensity of \Ha emission should be stronger than that of the other lines. 
Therefore, when our target is the \Ha line and the contaminant signal cannot be efficiently removed, it might be easier to estimate the \Ha 
line intensity from the total detected intensity than through power spectrum analysis.

Figure~\ref{fig:I_foreg_nu} also shows that, for $z_{{\rm H}\alpha}\lesssim 2$, OIII line emitters are the strongest line contaminants. 
The ratio between the intensity of the oxygen lines relative to that of the \Ha line emission is uncertain by a factor of about two at high redshift ($z\gtrsim5$).  
The uncertainty is due to the expected decrease in galaxy metallicity with increasing redshift. This contrasts with a few of the observed high redshift galaxies, where 
the line ratios of oxygen lines are actually quite high. Future observations with, for example, JWST will help to further constrain these lines intensities.

Another way of separating the different line contributions is by attributing a weight to the intensity of each line and using this 
information together with the frequency of the emitting lines to iteratively determine the line intensities. This algorithm would take advantage of 
the known/fixed separation between different lines and  directly fit the spectral information of all the lines 
intensities in a similar way to what was done for continuum infrared background data by \citet{2015Kogut}. Although, the accuracy of this 
method would be affected by the additional differential extinction suffered by each line.  

Moreover, the possibility of using the angular information of each of the lines in an intensity map was explored by 
\citet{2016Cheng} for the case of CII emission line. This study uses an MCMC approach to recover the intensity and bias 
information from each line. For the case of \Ha emission, however, this is a less promising approach, because there are several interloping lines originating from the same redshift. Moreover, the intensity of these lines is uncorrelated, whereas for CII the main contaminant lines, the different CO transitions, are highly correlated. This results in a much higher number of parameters for the case of the \Ha line and so the MCMC result would be very difficult to interpret. The solution for this problem would be to independently obtain strong constraints on some of the parameters and to properly determine the correlations between them prior to using the MCMC approach.   

\begin{figure}
\begin{centering}  
\includegraphics[angle=0,width=0.52\textwidth]{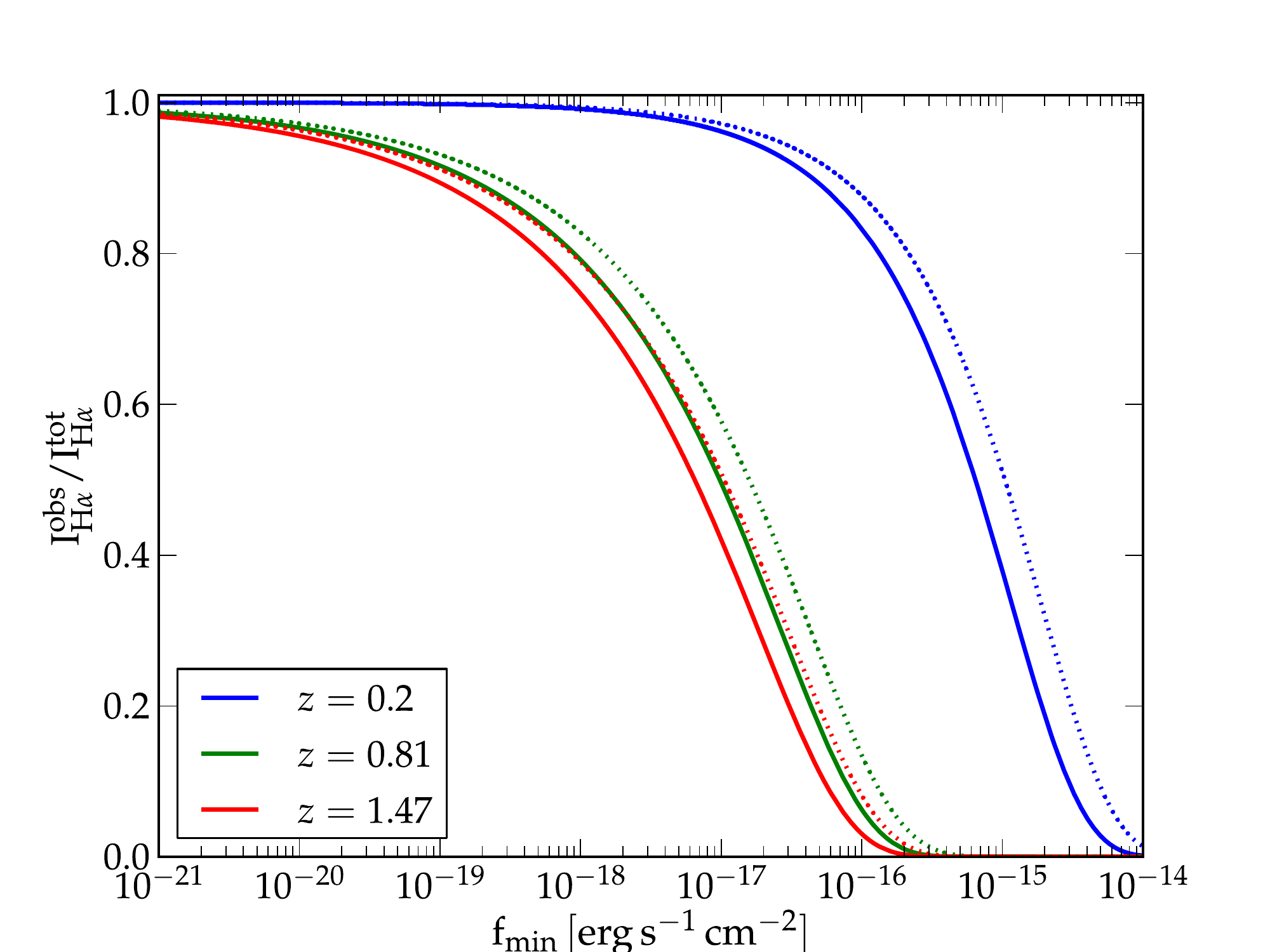}
\caption{Fraction of \Ha emission powered by star formation detected by a survey as a function of its flux sensitivity. Solid 
and dotted lines assume, respectively, an extinction in \Ha emission of $A_{{\rm H} \alpha}=1\,{\rm mag}$ and 
$A_{{\rm H} \alpha}=0.475\,{\rm mag}$. The lines shown correspond to redshifts $z=0.2$, $z=0.81$ and $z=1.47$, from right to left.  This 
figure is based on observational LFs from \citep{2015Stroe,2013Sobral,2016Smit}.}
\label{fig:I_Ha_fcut}
\end{centering}
\end{figure}

\begin{figure}
\begin{centering}  
\includegraphics[angle=0,width=0.52\textwidth]{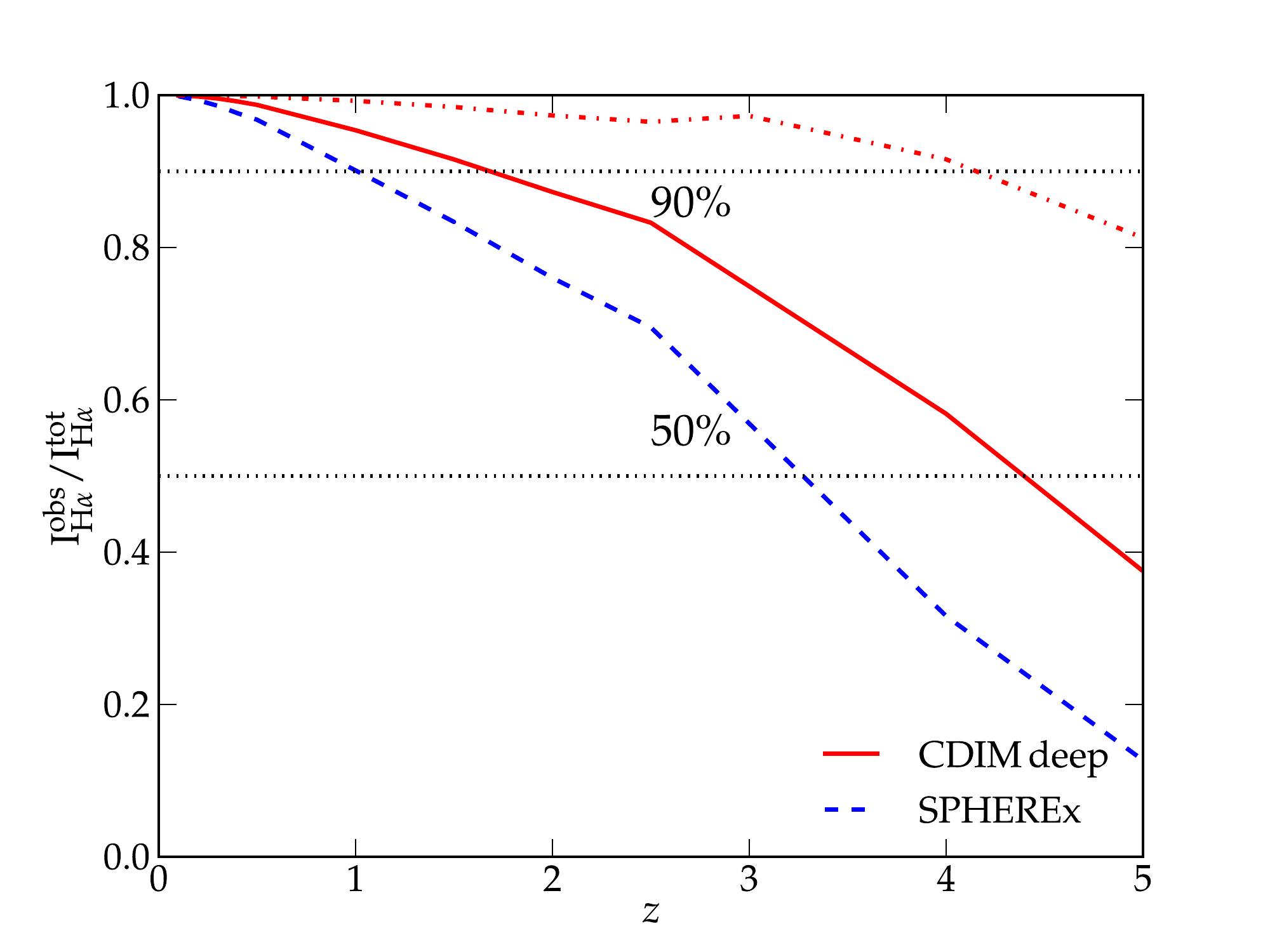}
\caption{Intensity of \Ha emission powered by star formation, detected by SPHEREx (blue dashed line ) and by CDIM (red lines). The 
red solid line assumes a flux limit of $4\times10^{-18}\, {\rm erg\, s^{-1}\, cm^{-2}}$, whereas the red dashed-dotted lines assumes a flux 
sensitivity of $1\times10^{-18}\, {\rm erg\, s^{-1}\, cm^{-2}}$.}
\label{fig:I_Ha_IM_perc}
\end{centering}
\end{figure}

Additionally, cross-correlating \Ha intensity maps with, for example, galaxy 
surveys of HI radio data, as was suggested by \citet{2017Gong}, can be used to avoid foreground contamination. Since, foregrounds contaminating \Ha emission 
and HI 21 cm data are uncorrelated to first order, the cross-correlation power spectrum would make it possible to probe the signal from these two lines. 
The contamination by higher-order correlations in this statistical measure is unfortunately not well explored yet. 
Also, this procedure is limited to 
the cases where an HI survey covering the same position in the sky and redshift range as the \Ha survey is available. 

\section{Astrophysical constraints from galaxy surveys versus IM surveys} 
\label{sec:Constraints}

While galaxy surveys can observe a small fraction of the Universe in great detail, IM surveys provide a global picture of our 
Universe by blindly detecting emission from all types of sources. 
IM surveys directly probe the global quantities, whereas we can only try to infer these quantities from the limited data provided by the 
selection biased and flux-limited galaxy surveys. 

In Figure~\ref{fig:I_Ha_fcut} we show the percentage of \Ha emission powered by star formation that can be probed by a survey as a 
function of its flux sensitivity.  This figure is appropriate to infer the flux sensitivity limits required for a galaxy survey 
to be able to detect a meaningful fraction of the \Ha emission originating from a given redshift. In Figure~\ref{fig:I_Ha_IM_perc} we show the intensity 
of \Ha emission powered by star formation that can be detected (assuming flux sensitivity limits, at least as high as the one indicated 
in Table~\ref{tab:instruments}) by the SPHEREx and CDIM instruments. We note that using a statistical analysis, such as the power spectrum analysis, would allow the 
detection of emission below these flux sensitivity cuts.
 
Throughout this study, we mainly focus on five central points that illustrate how each of the surveys described in Table~\ref{tab:instruments} is useful for 
astrophysical purposes and for specifically probing \Ha emission. These points are: the survey collecting area, its 
flux sensitivity, the ability to distinguish between emission powered by AGN or by star formation, the certainty in the identification of the emission line 
responsible for the observed flux and finally, the accuracy to which the dust extinction suffered by the observed line can be determined.

We now discuss how well these points are attained for galaxy surveys, and then for IM surveys.  

The properties of the Euclid and WFIRST surveys are similar, and for that reason, they will be able to detect the 
same types of galaxies. This study shows that, due to their sensitivity limits, these instruments will mainly detect bright galaxies. It also shows that,   
thanks to the planned large fields of view of these surveys, enough bright galaxies will be detected to further constrain the high luminosity end of the \Ha LF. 

Galaxies that are bright in \Ha emission can vary a lot in their properties, such as in their mass or the virial mass of the DM halo they 
belong to. Although  to a first approximation, massive 
galaxies are supposed to be the brightest, this will not always be true. These galaxies can suffer from quenched star 
formation, which would considerably decrease their \Ha emission. Also, massive galaxies are usually dusty galaxies and so their observed luminosity can be low. 
For these reasons, it is not necessarily true that the observed galaxies will correspond to a specific type of galaxy. Moreover, the observed bright 
galaxies might not be representative of the main \Ha galaxy population. As an example, bright galaxies might have 
particularly high or low luminosity ratios between different 
observational bands, compared to the majority of the \Ha emitters, which are much fainter.

Except for the small volumes magnified due to lensing, the spectroscopic surveys performed by WFIRST and Euclid will not 
reach the necessary low luminosities to probe the galaxies that dominate the overall \Ha line intensity. 
These surveys will also not probe the $\alpha$ slope (see Eq.~\ref{eq:LF}) of the \Ha luminosity function, since the lensed volumes 
are too small to beat cosmic variance. Furthermore, the uncertainty in the 
modelling of the lenses itself can be considerably high, especially for highly lensed sources \citep{2017Livermore}. These surveys might, however, help 
to identify the properties of a few relatively faint galaxies, given that the number density of these systems is high. Therefore, the lensed galaxies 
observed by Euclid and WFIRST will be important to understand the properties of low luminosity systems and therefore, to probe the evolution of 
the relation between \Ha emission and SFR in a galaxy.

For the WFIRST and Euclid planned surveys, the combination of the low-frequency range covered, the relatively broad photometric filters 
and the large number of sources that they are expected to detect, will make it impossible to properly probe the extinction suffered by each source. 
Line extinction will mainly be probed using galaxy templates. Although template-fitting algorithms will take into account dust 
extinction, it will not be possible to obtain accurate measurements for all sources, in particular for highly extincted sources \citep{2017Galametz}. 

On the other hand, these galaxy surveys will be particularly good at identifying AGNs and thus in distinguishing between star 
formation and AGN powered \Ha emission.

Moreover, Euclid and WFIRST are mainly being built with the objective of probing dark energy. For that propose, it is important to correctly 
identify the redshift of the source of emission.  The spectroscopic and photometric capabilities of these instruments will then, at 
least for sources with a high signal to noise ($S/N > 10$), allow them to 
determine which emission line is responsible for the observed signal \citep{2016Bisigello}. Ancillary data at lower wavelengths will also be used to help 
identify line contaminants in these surveys \citep{2016Bisigello}.

The case for IM surveys is very different. The advantage of the IM surveys performed by the SPHEREx and CDIM instruments 
resides in their low flux limits, combined with large frequency ranges and large FOVs. These factors will allow them to probe a large fraction 
of the intensity of \Ha emission over a large redshift range. This will also allow them to probe the time evolution of \Ha emitters.  
Moreover, by detecting emission from the sources mainly responsible for the total intensity of the \Ha line, their observational intensity maps 
can be used to probe the global star formation in these galaxies.

Constraining the SFRD with the SPHEREx and CDIM IM surveys will require updating the relation between \Ha emission and the SFRD. These updates should be made 
using constraints from other emission lines obtained with the same surveys, as discussed in Section~\ref{subsec:Z_ion_param}.
Moreover, the data from IM surveys will be difficult to interpret and separate in terms of the source of the emission.

\section{Summary and discussion} 
\label{sec:Summary}

In this study, we explored the potential of different instruments to constrain \Ha line emission. Namely, we compared the galaxy 
surveys that will be performed by the Euclid and WFIRST instruments, with the IM surveys that are planned for the SPHEREx and CDIM instruments.

Starting from observations, which we then extend by using physically motivated relations deduced from theory and simulations, we modeled 
the intensity and power spectra of the \Ha line over the $z\sim 0 - 5$ redshift range. 

We find that the intensity of this line is currently uncertain by a factor of up to a few until $z\sim 2$, and up to one order of magnitude at $z\sim5$. The higher uncertainty 
towards high redshift lies both in the lack of observations of \Ha emitters and in the increasing uncertainty of dust extinction corrections.
Still, the available constraints led us to estimate that this line intensity, in the relevant redshift interval 
should remain in the $I_{\rm H\alpha} \sim 10^{-8}-10^{-7}\, {\rm erg\, s^{-1}\, cm^{-2}\, sr^{-1}}$ range and peak at the same time as the cosmic SFRD at $z\sim 2-3$.

According to the properties of the considered CDIM and SPHEREx instruments, we predict that their planned IM surveys will be good enough to make a statistical detection of the overall \Ha line intensity. Moreover, when operating as a galaxy surveyor CDIM should detect more than $90\%$ of this line intensity up to $z\sim 4-5$, whilst SPHEREx can only do the same up to $z\sim1$. SPHEREx will still be able to detect more than $50\%$ of this emission up to $z\sim4$. These percentages assume a minimum flux per observational voxel corresponding to the flux limits quoted for these surveys. 

On the other hand, the Euclid and WFIRST galaxy surveys will only detect \Ha emission in a narrow frequency range. This will include \Ha emission only up to $z\sim2$. Given 
their flux sensitivity limits, in spectroscopic mode, these surveys will not probe the \Ha line intensity.
However, they will probe the high end of the \Ha LF over a large enough volume to beat cosmic variance. Moreover, at least for bright sources, Euclid and WFIRST will   
be able to distinguish between SF and AGN powered \Ha emission. 
The photometric capabilities of these surveys will also be used to probe the galaxy dust extinction and 
to help distinguish between emission from different lines. These two points will only be easily achievable for relatively bright galaxies.

Using the same methodology as for the \Ha line, we modeled the emission by the SII and NII doublet lines and by the OII, OIII, H$\beta$ and Ly$\alpha$ lines 
in the redshift range where these lines will contaminate \Ha intensity maps. We found that for a survey not suffering from flux limitations, 
the signal from contaminant lines will increase the observed \Ha power spectra  by a factor up to two at $z\lesssim 2$. At higher 
redshift this contamination is expected to decrease.     

We implemented the several models for line emission in a simulation code and obtained a light cone for both the \Ha line and the background 
contaminant lines. The observational light cones assume the flux sensitivity and voxel resolution of each of the planned IM surveys. We find that, 
besides the contamination by background interloping lines being quite strong  (both in terms of intensity and power spectrum), it is also difficult to remove. 

We applied flux cuts to the background lines of $(3.0,~1.2,~0.5)\times 10^{-16}~{\rm ergs\, s^{-1}\, cm^{-2}}$, in order to estimate the effect 
that a masking procedure would have on the contamination power spectrum. 
We find that the required flux cuts for masking need to be stronger towards increasing redshift. Still, overall they were successful at decreasing 
the contamination power spectrum to a maximum of $\sim 10\%$ of the observed \Ha signal. The removal of contamination by these bright background 
galaxies would only require masking less than $1\%$ of the voxels for both the SPHEREx and the CDIM surveys.
The decrease in the \Ha power spectrum due to putting the strongly contaminated voxels to zero would also be below $1\%$. The 
recovery of the target signal would thus not be compromised. 

At $z\lesssim 0.8$, the detection of interloping contaminants in \Ha intensity maps can be
reasonably done with data from a galaxy survey with a flux sensitivity of $1.2\times 10^{-16}\, {\rm ergs\, s^{-1}cm^{-2}}$. This corresponds 
to the sensitivity of the WFIRST instrument, although it lies outside the frequency range covered by this survey. For \Ha intensity maps 
in the range $z\sim0.8-2$, a flux cut of $5.0\times 10^{-17}\, {\rm ergs\, s^{-1}cm^{-2}}$ would produce a similar reduction of the contaminants power spectrum.
Unfortunately, there is no available galaxy survey that can detect contaminant galaxies up to this low flux level and over the large volumes covered by the IM surveys. 

Given the lack of surveys that can individually detect the main contaminant galaxies, we believe that the best option for 
\Ha IM studies is to jointly model the evolution of all the strong emission lines contributing to the observed line fluxes. After recovering the observed \Ha signal, intensity maps should be corrected for dust extinction. We explored the possibility of 
doing this with the data from the same IM surveys and found that this can be done, to a certain point, using ratios of emission lines. 
This would require efficient separation of the contribution from the different lines in these intensity maps, which is not trivial. 
Alternatively, the extinction rates and extinction curves from galaxy surveys can be used to at least predict the overall evolution 
of the dust extinction rate at the relevant frequencies.   

We, therefore, conclude that IM surveys can be used to probe the overall intrinsic \Ha intensity up to an uncertainty of the order of $20\%$, as long 
as this line signal can be accurately disentangled from the signal of other emission lines.
 
Many of our results are biased towards our choice of using a high redshift SFRD model based on observations of \Ha emitters. These have 
inferred SFRs higher than the SFRs traced by the UV continuum. Despite the reasons for this discrepancy, our choice allowed us to make 
predictions for the \Ha intensity and power spectrum consistent with observations. Moreover, OIII emitters, which are the most important contaminants 
in \Ha intensity maps, also have very large equivalent widths at high z. Consequently, our SFRD based predictions for the intensity of this 
line are consistent with observational constraints.
In the case of OIII emitters the large EWs are likely to be the result of the high ionized state of the ISM in high redshift galaxies. 
Therefore, we might be overestimating the contamination by H$\beta$ and OII emitters. However, this does not affect our main 
conclusions, given the small contribution from these lines ($\lesssim10\%$) to the contamination power spectra in \Ha intensity maps. 

\section*{acknowledgements}
The authors thank the anonymous referee whose comments and suggestions helped to improve the quality of the article.
We also thank the Netherlands Foundation for Scientific Research support through the VICI grant 639.043.006.

\bibliography{Ha}

\appendix
\section{Constraints from IM and galaxy surveys}\label{app:constraints}

\subsection{Constraints on the properties of \Ha line emitters}
\label{subsec:Const_Ha_obs}

In this appendix, we discuss, in some detail, different constraints on galaxy properties that can be achieved by galaxy and IM surveys targeting \Ha emission.

In order to properly constrain \Ha emission, a survey needs to cover a large volume to overcome cosmic variance. It also has to be sensitive 
enough to detect emission from relatively faint sources in order to probe a large fraction of the overall intensity of line emission.
Moreover, it should cover a large bandwidth in frequency to probe the target signal over a 
large redshift/time range and to allow for the identification and removal of line contaminants.
Spatial and spectral resolution is also important if one wishes to identify the individual sources of emission, or at least to 
differentiate between star-forming systems and active galactic nuclei. Furthermore, the ability 
to determine the extinction suffered by the \Ha line is needed to probe this line intrinsic luminosity.

We now describe, in some detail, how each of the surveys described in Section~\ref{sec:Surveys} can satisfy the above-listed requirements to constrain \Ha emission.

\subsubsection{Cosmic Variance}
Euclid and WFIRST aim to detect and characterize \Ha emission from high luminosity galaxies.
Massive, luminous galaxies are preferentially located in the high-density peaks and so these instruments are 
required to survey a large fraction of the sky in order to beat cosmic variance. 
This implies covering volumes of the order of $5\times 10^5\, {\rm Mpc^3}$, just to beat cosmic variance in \Ha emission 
at the 90\% level \citep{2015Sobral}. 

The combination of Euclid's deep survey frequency resolution and its field of view (FOV) translates to a volume per 
redshift bin too small to beat cosmic variance. As a result, Euclid's  data of bright galaxies will have to be averaged in three or four 
frequency bins before being used for \Ha studies. 
This will not be necessary for WFIRST data due to its impressive 2227 ${\rm deg^2}$ FOV. 
Both the SPHEREx deep survey and CDIM wide survey will cover large enough volumes to beat cosmic variance.

\subsubsection{Sensitivity} 

The observed characteristic \Ha luminosity, $L^{\star}_{\rm H\alpha}$, scales as ${\rm log}\, L^{\star}_{\rm H\alpha} [{\rm erg\, s^{-1}}]= 0.45z + 41.47$ 
over the redshift range $z=0-2.23$ \citep{2016Sobral.Kohn}.

The Euclid and WFIRST spectroscopic surveys will be able to constrain the \Ha LF above the characteristic luminosity at $z\sim 0.7$ and $z\sim1.1$, 
respectively.
However, since galaxies below the $L^{\star}_{\rm H\alpha}$ luminosity still contribute significantly to this line LF, these instruments will 
not have enough sensitivity to constrain the overall \Ha intensity. Note that their photometric surveys will be much better at constraining this intensity.  

In the case of IM surveys performed by the CDIM and the SPHEREx instruments, the lines intensities can be totally recovered using statistical methods. Moreover, the signal in bright voxels of the observed maps, can be independently detected as is done with traditional galaxy survey data. In that case, only the voxels with a signal above the noise can be recovered, where the noise in the map is given by the flux sensitivity of the IM survey listed in Table~\ref{tab:instruments}.  

The surveys planned for CDIM and SPHEREx will have better flux sensitivities than those performed by Euclid or WFIRST. Therefore, these 
IM surveys should be able to directly probe the \Ha intensity in most voxels at low $z$ and a reasonable fraction of this signal at higher $z$. CDIM can do much 
better than SPHEREx on this point, as is shown in Figure~\ref{fig:I_Ha_IM_perc}. This figure shows that in CDIM intensity maps, the \Ha signal should be above the noise level in close to 80\% of the voxels at $z\sim5$. Moreover, this percentage will increase towards lower redshifts.

\subsubsection{Contamination} 
\label{app:Contamination}
\textbf{AGN contamination:} AGN mostly populate the high end of the \Ha LF.
The number density of AGN, contributing to the \Ha LF, at $z<2.3$ is fairly
constant \citep{2016Sobral.Kohn}. Given, the expected evolution of star forming \Ha emitters, the AGN fractional contribution to the overall \Ha intensity is expected to increase from $z = 0$ towards the peak of star formation activity. However, the \Ha line intensity will remain dominated by fainter star-forming galaxies \citep{2016Sobral.Kohn}. 
Moreover, the bright AGN population should rapidly decline at $z\simeq3$, which  will 
consequently decrease its contribution to the \Ha intensity \citep{2015Haardt.Salvaterra}.

By integrating over the \Ha luminosity function and assuming that the AGN population corresponds to 30\% of 
the observed \Ha emitters \citep[see][]{2016Sobral.Kohn} the derived number density of bright AGN is of the order of $10^{-4}\, {\rm Mpc^{-3}}$.

For very bright sources, Euclid and WFIRST surveys will be sensitive enough to differentiate between \Ha emission powered by AGN activity or by star formation. 
In these surveys, AGN will be identified as sources with Balmer lines with FWHM above ${\rm 1000\, km\, s^{-1}}$. When possible, AGN will also be identified 
using emission line flux ratios \citep[see][]{2016Sobral.Kohn}.

In IM surveys data, the sources of \Ha photons are completely blended together. Nevertheless, thanks to the increasing number of AGN detected at high-z, the brightest AGN can be masked from intensity maps significantly reducing their signal.

The voxels 
from the SPHEREx surveys have volumes of the order of a few ${\rm Mpc^3}$ whereas the voxels from CDIM surveys are even smaller. Therefore, 
contamination by bright AGN should be present in a small percentage of the voxels of the observed intensity maps.

There will not be enough information about AGN at high redshifts to mask all these galaxies from intensity maps. 
However, their overall contribution to the \Ha intensity, as predicted by integrating over current \Ha LFs, should be below the 1\% level. 

The identification of AGN in the presence of multiwavelength data is further discussed in Section~\ref{sec:App.AGN}. 

\textbf{Line contamination:} \Ha observations will suffer strong contamination from several background lines.  
Surveys covering a large frequency range with a relatively high spectral resolution are necessary to allow for an accurate identification of the 
emission line responsible for the observed signal. 

Euclid and WFIRST will obtain low-resolution spectra of the energy distribution of 
resolved galaxies emission. This will, in some cases, allow for the identification of the targeted emission line. At this point, 
Euclid will have the advantage over WFIRST, given its higher spectral resolution.

IM surveys do not resolve emission sources. For these surveys, additional, independent surveys are used to identify bright interloping lines. 
Alternatively,, the signal of several emission lines can be fitted to power spectra derived from the intensity maps at different redshifts. This can potentially be used to separate the emission in different emission lines.

\subsubsection{Dust extinction} 
\label{sec:App.Dust}
Both galaxy surveys and IM surveys will be sensitive to the \Ha radiation that escaped the galaxy without being absorbed by the dust in the ISM. However, for astrophysical 
studies, the relevant quantity is the intrinsic \Ha emission and so the observed \Ha emission has to be corrected for dust extinction. 

Dust extinction is usually measured by fitting a galaxy emission spectra 
and/or with observed ratios of emission lines. 
Most commonly, \Ha line extinction in the interstellar medium is estimated 
from the observed fluxes in the Balmer \Ha and H$\beta$ emission lines. The emission rates of recombination and collisional excitation of these lines scale approximately in 
the same way with the gas temperature and ionization state. Therefore, the intrinsic ratio between these lines fluxes is expected to follow the ratio $f^i_{\rm H_{\beta}}/f^i_{\rm H_{\alpha}}=0.35$, as mentioned 
in Section~\ref{sec:contamination1}. With this method, the magnitude of extinction suffered by the \Ha line is:
\be
A_{\rm H\alpha} = \frac{-2.5\, k_{\rm H\alpha}}{k_{\rm H\beta}-k_{\rm H\alpha}}{\rm log_{10}}\, \frac{f^i_{\rm H\alpha}/f^i_{\rm H\beta}}{f_{\rm H\alpha}/f_{\rm H\beta}},
\ee
where $f_{\rm H\alpha}$ and $f_{\rm H\beta}$ refer, respectively, to the observed line fluxes in the \Ha and H$_{\beta}$ lines and the index i 
denotes the intrinsic line fluxes \citep{2012Sobral}. We assume the dust absorption coefficients, $k_{\rm H\alpha}$ and $k_{\rm \beta}$, given by the \cite{2000Calzetti} extinction curve.
The \Ha dust extinction is therefore
\be
A_{\rm H\alpha} = 6.531\, {\rm log_{10}}\, f_{\rm H\alpha}/f_{\rm H\beta}\, -\, 2.981.
\label{eq:A_alpha_beta}
\ee

This equation will be used to estimate the dust extinction suffered by bright galaxies in the galaxy surveys data. Euclid will 
detect both the \Ha and \Hb fluxes, for \Ha emitters in the redshift range $z\sim1.29-2.04$. While WFIRST
will only be able to do the same in the much smaller $z\sim1.8-1.96$ redshift range.
Euclid and WFIRST spectroscopic surveys will cover frequency ranges to narrow to allow to properly fit a galaxy spectra. Therefore, when no other constraints are available, these missions will rely on dust extinction rates estimated from the relative intensities in their B and V filters \citep{2017Galametz}.  

The relatively large frequency range covered by IM surveys allows for the use of several different methods to probe the overall \Ha
dust extinction. For these surveys, dust extinction will be estimated using:  
flux ratios of high transition hydrogen lines, flux ratios of metal 
lines, by fitting low-resolution spectra and by taking advantage of extinction rates derived from other missions.

At the local Universe, SPHEREx will map the entire sky, with Paschen-$\alpha$ ${\rm \left( 1875\, nm\right)}$ and Brackett-$\alpha$ 
${\rm \left( 4051\, nm \right)}$ recombination lines. These lines are intrinsically weaker than the \Ha line, but 
they suffer less dust extinction than the Balmer \Ha and \Hb lines and so they are better probes of galaxy properties. 
At $z\, \lesssim\, 0.5$, the amount of dust extinction will 
be probed using features in low-resolution spectra of millions of galaxies per bin. At $z\sim1$, this survey will make use of photometry 
from LSST and other surveys in order to probe the dust content of the galaxies. Finally, at $z~\sim~1.36$, SPHEREx will be able to use the 
H$\beta$ plus the OIII [${\rm 500.7\, nm}$] lines to estimate the dust extinction in the galaxy. This data will also be compared 
to galaxy SEDs derived from previous surveys in the same redshift ranges, as described in \citep{2016Dore}

In order to probe for dust extinction, the proposed CDIM mission will have 
all the resources available for SPHEREx. Additionally, CDIM will have a higher frequency resolution which allows for the separation between emission in very close lines. This is 
essential for the SED fitting technique to be successful. Therefore, CDIM will provide more reliable estimates of \Ha dust extinction than SPHEREx. 

Since there is no observational basis indicating that dust extinction curves evolve considerably up to $z\sim5$, the main source of uncertainty in 
the extinction magnitudes derived from observed line flux ratios, involving metal lines, are the assumed intrinsic flux ratios. 

The intrinsic fluxes, of hydrogen lines, scale mainly with the intensity of the ionizing radiation, which is one of the galaxy properties more easily accessible to observations. On the other hand, the fluxes of metal lines also depend strongly on additional galaxy properties, such as gas clumping, metallicity and hardness of the ionizing spectrum. The redshift evolution of these additional galaxy properties is not easy to infer from observations, as discussed in Section \ref{subsec:Z_ion_param}.

At redshift windows where an estimation of the dust extinction cannot be made, the intrinsic \Ha emission cannot be recovered up to a $20-40$\% correction factor.

\subsection{Constraints on ISM gas metallicity and ionization state}
\label{subsec:Z_ion_param}

The Universe at high redshifts was considerably younger and so galaxy properties, such as IMF, metallicity and SED 
might have been very different from today's galaxies. 

High-z observations are often interpreted using relations based on observations at very low-z. 
The evolution of galaxy properties will thus affects the 
reliability of the interpretation of high-z observational data.
This problem can be mitigated using multiwavelength sets of  data to confirm/constrain key properties of high-z galaxies. 

Current constraints indicate that, although galaxy metallicity is a function of stellar mass and SFR, this dependence shows little to no evolution up to 
$z\sim3.7$ \citep{2016Hunt.Dayal}. At higher z, there are no available observations that can confirm or refute this trend. 
Towards very high-z ($z\gtrsim 6$), however, galaxies are expected to be more metal-poor, due to the low amount of time available for the stars to pollute the ISM with metals, and so the intrinsic ratios of metal line fluxes to the galaxy SFR should decrease. 

The dust-to-gas mass ratio in the interstellar medium scales with metallicity since a substantial fraction of heavy elements is part of the dust composition.
Simulations predict that the average dust mass density in the ISM declines by around one order of magnitude from $z\sim0$ to $z\sim5$ \citep{2016Popping}. 
On the observational side, infrared emission by dust seems to be consistent with no evolution in the ISM dust mass up to a redshift of $z\sim2.3$ \citep{2013Thacker}.
In the near future, observations of individual galaxies by the ALMA and JWST instruments should confirm or refute these trends.
 
Galaxy surveys can probe the metallicity of galaxies through their emission spectra or with ratios of strong metal lines, namely carbon, oxygen and nitrogen lines. 
The Euclid and WFIRST surveys will provide low-resolution spectra that can only be used to constrain the properties of very bright galaxies. They will also probe galaxy metallicity and ionization state using the EW of the OIII emission line, which will be detected in the redshift range $\left[ 1.2/1.7, 3.0\right]$ for, respectively, the Euclid and WFIRST surveys. These surveys will span a narrow frequency 
range which will not allow them to detect many metal lines in overlapping redshift ranges. Therefore, the Euclid and WFIRST missions plan to make use of data from other instruments/surveys to probe more frequency bands and so to further constrain their galaxy spectra. 

In the case of IM surveys, the 
relative intensity of emission lines can be used for this same objective. Due to their large frequency range, the described IM surveys will be able to detect the integrated emission from the H$\alpha$, H$\beta,$ 
OII and OIII lines at $z\sim0.87-5$. 

In intensity maps, the signal from different lines will be blended together. However, there are frequency 
windows where a specific line dominates the total intensity in an observational voxel \citep{2017Fonseca}. This can be used to help 
disentangle the contribution from each line so that their intensities can at least be fitted as a function of redshift. 
In that case, these line intensities can be used to constrain the overall galaxy metallicity and ionization parameter. A method to do this is described in detail in Section~\ref{sec:App.B}.

\subsection{Constraints on the SFRD}
\label{subsec:Const_SFRD}

Prior to being used in SFR studies, the observed \Ha signal needs to be corrected for line contamination, AGN emission, and dust extinction as discussed, 
respectively, in Section~\ref{sec:LineContam} and in Section~\ref{subsec:Const_Ha_obs}. Moreover, in order to constrain the SFRD the \Ha luminosity density needs to be probed, which implies observing down to relatively faint levels. 

As is clear from Figure \ref{fig:I_Ha_fcut}, the spectroscopic surveys performed by Euclid and WFIRST will 
not have enough sensitivity to probe the overall \Ha luminosity density. As a result, on their own, these 
surveys will not be able to constrain the cosmic SFRD. They will, however, be able to individually detect 
and identify many AGN.  
Euclid and WFIRST will use their photometric surveys to detect fainter \Ha emitters, but not faint enough to constrain the overall \Ha emission. These instruments photometry will also be helpful in estimating the dust 
extinction suffered by the \Ha line. 

IM surveys, should be able to make a statistical detection of the full \Ha signal. The observed signal will be contaminated by several interloping lines. However, either by masking the contaminated voxels or by fitting the emission from the several lines, the \Ha intensity as a function of redshift should be recovered 
with reasonable accuracy, as discussed in Section~\ref{sec:LineContam}. The dust extinction suffered by this
line can also be estimated using the procedures described in Section~\ref{sec:App.Dust}. Moreover, the contribution from AGN, to the total \Ha intensity, is relatively small and in principle the brightest AGN can be masked in the observational maps, as discussed in Section~ \ref{app:Contamination}.

IM surveys will, therefore, be able to constrain the intrinsic \Ha intensity reasonably well up to $z\sim 5$. 

The SFR can be inferred from the intrinsic \Ha luminosity using Equation~\ref{eq:Lum_Ha}. The physics connecting \Ha emission and SFR sets the coefficient in this equation. At low redshift, there are enough observations 
to constrain this coefficient, however, that is not the case at high-z ($z\gtrsim2$).
The validity and accuracy of this coefficient depends on the following assumptions: 
 
\begin{enumerate}

\item
\label{item}
The gas is in ionization equilibrium at a temperature of $T={\rm~10^4}~{\rm K}$ and \Ha photons are mainly emitted during recombinations. 
Detailed simulations of the ISM and comparison with observations indicate that, at solar metallicity,
collisional excitations contribute to about $\sim\, 1\, -\, 10\%$ of the total \Ha emission \citep{2017Peters}. 
Higher contributions would require very strong UV fields that would destroy molecular clouds and cause a subsequent decline in the galaxy star formation rate.  
There are not many observational constraints at high redshift and so collisional excitations cannot be safely ignored if we allow the gas to be at a higher temperature, as is shown in Figure~\ref{fig:coll_rec}.
 
\item
\label{2item}
All UV ionizing photons are spent on ionizing the ISM gas. This gas will eventually recombine, emitting several line photons. 
This premise assumes that the escape fraction of ionizing UV radiation is very close to zero, which is consistent with the small number of observational constraints
at low redshift. On the other hand, a larger escape fraction is expected towards higher redshifts in order to account for the intensity of the UVB radiation and to explain the reionization of hydrogen gas.
  
\item
\label{3item}
The probability of emission of an \Ha photon per hydrogen recombination is $f_{\rm rec}=0.45$. In reality, this factor will depend on the galaxy IMF and on the strength of the UV field. At the local Universe, the assumed $f_{\rm rec}$ value is consistent with observations, but towards high redshifts, there is a discrepancy between SFRDs derived from \Ha fluxes and from continuum UV luminosities \citep{2016Smit}.
This discrepancy might be explained with bursty star formation episodes, boosting the \Ha emission (hypothesis discouraged by observations). It can also be explained by a particularly 
strong UV radiation, which would be a consequence of changes in the IMF and/or in the UV $\beta$ slope (supported by observations). An alternative explanation would be that the dust extinction curve assumed is wrong and consequently  
the relation between the extinction in the two frequency bands would be different from what is assumed. 
The future JWST will help to identify the source of this discrepancy and confirm if there is any meaningful redshift evolution in the amount of dust or in the galaxy spectral energy distribution.  

\end{enumerate}
 
Despite the several issues pointed out above, the coefficient between \Ha emission and the SFR can be adjusted using future observational data. As a result \Ha luminosity density can be used a good tracer of the global SFRD even at high redshifts.

\subsection{Constraints on the stellar mass}
\label{subsec:Const_mstar}

The most reliable way to estimate a galaxy stellar mass is likely to be fitting the galaxy emission spectra to theoretical templates of galaxies SED. However, this method will only work when the galaxy SEDs is well constrained and even then its accuracy depends on the validity of the templates used. 

There are multiple galaxy properties that have a similar impact on a galaxy SED. For example, the age of a galaxy, metallicity and dust extinction will all affect the same frequency range corresponding to the red band. Note that towards high redshifts, galaxies will be all relatively younger (due to the young age of the Universe) and dust extinction and metallicity will be smaller, making the stellar mass estimates more robust. 

Alternatively, the stellar mass of a galaxy can be inferred from ratios of fluxes in two frequency band or from the intensity of some emission lines. This method has the advantage that the degeneracy between many of the parameters that shape a galaxy SED can be broken or minimized by using ratios between emission lines.

The connection between galaxies stellar mass and ratios between fluxes in different bands or the strength of an emission line is biased towards the chosen IMF. By default, in studies of the high redshift Universe, it is assumed 
that galaxies follow a Salpeter IMF. A heavier IMF would result in a smaller stellar mass as inferred from line emission data. 
Moreover, a galaxy SED affects extinction curve of a galaxy. Prescriptions to improve upon extinction estimates suffered by an emission lines, when the galaxy SED is unknown, are discussed in \citet{2017Galametz}.

For Euclid and WFIRST, bright galaxies stellar mass will be estimated by fitting the observed low-resolution spectra to theoretical templates.
The wavelength range covered by Euclid and WFIRST corresponds to the R band, where extinction is not as high as in the UV band. It is still an important issue, however.
The photometric capabilities of these missions will allow for the estimation of the galaxy metallicity through the diagnostic $R_{23}$ ratio given by:
\be
{\rm log\, R}_{\rm 23} \equiv \left( \frac{I_{[{\rm OII}]\lambda3727}+I_{[{\rm OIII}]\lambda4959}+I_{[{\rm OIII}]\lambda5007}}{I_{{\rm H}\beta}} \right).
\label{eq:R23}
\ee
These estimates will be somewhat unreliable, given that the frequency band covered by each filter is very large. Therefore, a filter will collect emission from several transition lines. Also, the redshift 
range for which the estimate of the $R_{23}$ ratio is possible is small.

IM surveys, will not be able to measure galaxies spectra. However, the SPHEREx and CDIM surveys will reach near-infrared wavelengths, where dust extinction is relatively small, and so these instruments surveys can potentially provide good estimates of galaxy stellar mass.

At low redshift, SPHEREx will map galaxy emission in the Paschen and Bracket hydrogen recombination lines. These are infrared lines and so they suffer from 
very low extinction, which makes them relatively good probes of galaxy stellar mass and star formation rate. The sensitivity of SPHEREx will only allow it 
to probe these emission lines in bright galaxies, down to a stellar mass of $\sim 10^{10}\, {\rm M}_{\odot}$. This lower limit on the stellar mass is high 
compared to the stellar mass range that is currently probed by other instruments targeting \Ha emission \citep{2011Villar,2016Sobral.Stroe}.
On the other hand, CDIM has a higher sensitivity than SPHEREx, by about one order of magnitude, and so it should be able to detect galaxies with stellar masses as low as the current high-z galaxy surveys. 

In simulations such as Illustris, a tight correlation between star formation and stellar mass has been found at a fixed redshift 
\citep{2015Sparre}. Under this assumption, stellar mass can be estimated from the SFR derived from line emission observations. However, given the lack of observational constraints on the high-z ratio between star formation and stellar mass, it is advisable to use several line probes to confirm this relation. This is possible since some line strengths scale with the galaxy instantaneous SFR (such as the \Ha line), whereas others are proportional to the galaxy gas content or its stellar mass. 

\section{Multiwavelength data}\label{app:Multiwavelength}

In the presence of multiwavelength data, the constraints from high-z observations will become a lot more reliable. Therefore, all the missions discussed in this study will try to use all their available data to improve/confirm their conclusions. In this section we present some strategies involving multiwavelength data relevant for these missions. Moreover, we discuss some of the limitations associated with these methods. 

\subsection{Identifying AGN}
\label{sec:App.AGN}
The Baldwin-Phillips-Terlevich (BPT) diagram can be used to distinguish AGN from purely star-forming galaxies using the OIII/\Hb and NII/\Ha line ratios. This diagram has been updated by \citet{2013Kewley.Maier}, in order to be compatible with the redshift evolution of the ISM of star-forming galaxies. There are, however, selection biases associated with the BPT diagram, that cause some AGN in low mass and/or in high specific star formation rate galaxies to be missed \citep{2015Coil}. 
This is increasingly problematic towards high redshift.

\subsection{Constraining galaxies ISM}
\label{sec:App.B}
In the presence of fluxes from the H$\alpha$, H$\beta,$ 
OII and OIII lines, many ISM properties can be determined by iteratively converging the following set of equations, taken from \cite{2004Kobulnicky}.

The relative intensity on many lines depends on the ionization state of the emitting gas which can be fairly characterized by the ionization parameter (q).
The ionization parameter is given by $q= 10^{\rm A/B}$, with
\ba
A &=& 32.81-1.153y^2+\left[{\rm 12+log(O/H)}\right]\\ \nonumber
&\times& (-3.396-0.025y+0.1444y^2)
\ea
and 
\ba
B &=& 4.603-0.3119y-0.163y^2\\ \nonumber 
&+& \left[{\rm 12+log(O/H)}\right]\\ \nonumber 
&\times& (-0.48+0.0271y+0.02037y^2).
\ea
The parameter $y$ is the logarithm of the $O_{32}$ ratio between oxygen lines:
\be
y\equiv {\rm log} \left(O_{32}\right)\equiv {\rm log} \left( \frac{I_{\rm [OIII]\lambda4959}+I_{\rm [OIII]\lambda5007}}{I_{\rm [OII]\lambda3727}}  \right),
\ee
which is a function of the gas temperature and of its electron density.

The ISM metallicity can be probed with the diagnostic ratio $R_{\rm 23}$ given by Equation~\ref{eq:R23}.
Using $x\equiv {\rm log\, R}_{\rm 23}$, we have:
\ba
{\rm 12+log(O/H)} &=& 9.4+4.65x-3.17x^2-{\rm log(q)}\\ \nonumber
&\times& \left( 0.272+0.547x-0.513x^2\right)
\ea
for $\left( {\rm 12+log(O/H)\, <\, 8.4}\right)$ and
\ba
{\rm 12+log(O/H)} &=& 9.72-0.777x-0.951x^2\\ \nonumber
&-&0.072x^3 - 0.811x^4-{\rm log(q)}\\ \nonumber
&\times& ( 0.0737-0.0713x - 0.141x^2\\ \nonumber
&+& 0.0373x^3-0.058x^4)
\ea
for $\left( {\rm 12+log(O/H) \gtrsim8.4}\right)$.\\

An improved estimate of galaxy metallicity can be obtained with a fit from \citet{2016Dopita.Kewley}, given by:
\ba
12+ {\rm log(O/H)} &=& 8.77+ {\rm log} \frac{I_{\rm [NII] \lambda 6584}}{\rm I_{\rm [SII]\lambda\lambda 6717.31}}\\ 
&+& 0.246 {\rm log} \frac{I_{\rm [NII] \lambda 6584}}{I_{\rm H\alpha}}.
\nonumber
\ea 
This fit has the advantage of depending only on lines with similar frequencies.  

The gas metallicity and the $R_{\rm 23}$ ratio can be used to probe the galaxy dust content and, more specifically, the 
extinction power at the \Ha frequency. 
Also, the constraints on the SFRD derived from \Ha observations are improved by the extra estimate of the 
strength of the UV field obtained from the ionization parameter and the $O_{32}$ ratio.

Note that the ratio between the H$\beta$ $({\rm 468.8\, nm})$ and the OIII $({\rm 500.7\, nm})$ lines, used in the estimation of the R23 ratio. should evolve strongly with redshift.
At solar metallicity, the OIII line is intrinsically stronger than the H$\beta$ line. However, like the overall metallicity, the OIII line strength relative to the \Hb line is expected to scale strongly with galaxy mass and to decrease with increasing redshift. Moreover, the amount of doubly ionized oxygen depends on the ionization state of the gas in HII regions. This, in turn, is a function of the density/clumpiness of this gas \citep{2013Kewley.Dopita}.

\bsp	
\label{lastpage}
\end{document}